\documentclass[usenatbib]{mn2e}
\usepackage{graphicx} 
\usepackage{abbreviations} 
\usepackage{natbib} 
\usepackage{amssymb}

\newcommand{\be}{\begin{equation}}
\newcommand{\ee}{\end{equation}}
\newcommand{\oder}[2]{\frac{d #1}{d #2}}

\newcommand{\fracp}[2]{\left(\frac{#1}{#2}\right)}

\newcommand{\beq}{\begin{equation}}
\newcommand{\eeq}{\end{equation}}

\newcommand{\sub}[1]{_{\mbox{#1}}}

\graphicspath{{figures/}}

\begin{document}

\title[ Stability of cosmic jets]
{Causality and stability of cosmic jets}

\author[Porth \& Komissarov] 
{
Oliver Porth$^{1,2}$\thanks{E-mail: o.porth@leeds.ac.uk (OP)},
Serguei S. Komissarov$^{1,3}$\thanks{E-mail: s.s.komissarov@leeds.ac.uk (SSK)}\\
$^{1}$Department of Applied Mathematics, The University of Leeds, Leeds, LS2 9JT, UK \\
$^{2}$Centre for mathematical Plasma Astrophysics, Department of Mathematics, KU Leuven,
 Celestijnenlaan 200B, 3001 Leuven, Belgium \\ 
$^{3}$Department of Physics and Astronomy, Purdue University, West Lafayette, 47907-2036, USA}

\date{Received/Accepted}
\maketitle
\begin{abstract} 
In stark contrast to their laboratory and terrestrial counterparts, cosmic jets 
appear to be very stable. They are able to penetrate vast spaces, which exceed by up 
to a billion times the size of their central engines. We propose that the reason 
behind this remarkable property is the loss of causal connectivity across these jets, 
caused by their rapid expansion in response to fast decline of external pressure with 
the distance from the ``jet engine''. In atmospheres with power-law pressure distribution, 
$p_{ext}\propto z^{-\kappa}$, the total loss of causal connectivity occurs, when
$\kappa>2$ -- the steepness which is expected to be quite common for many astrophysical 
environments. This conclusion does not seem to depend on the physical nature of 
jets -- it applies both to relativistic and non-relativistic flows, both 
magnetically-dominated and unmagnetised jets.     
In order to verify it, we have carried out numerical simulations 
of moderately magnetised and moderately relativistic jets. The results give strong 
support to our hypothesis and provide with valuable insights.  
In particular, we find that the z-pinched inner cores of magnetic jets expand slower 
than their envelopes and become susceptible to instabilities even when the whole jet is 
stable. This may result in local dissipation and emission without global disintegration 
of the flow. Cosmic jets may become globally unstable when they enter flat sections of 
external atmospheres.  We propose that the Fanaroff-Riley morphological division of extragalactic 
radio sources into two classes is related to this issue. In particular, we 
argue that the low power FR-I jets become re-confined, causally connected and globally 
unstable on the scale of galactic X-ray coronas, whereas more powerful FR-II jets 
re-confine much further out, already on the scale of radio lobes, and remain largely 
intact until they terminate at hot spots. Using this idea, we derived the relationship between 
the critical jet power and the optical luminosity of the host galaxy, which is in a very good 
agreement with the observations.
        
\end{abstract}


\section{Introduction}
\label{sec:Intr}

Collimated outflows (jets) from stars and active galactic nuclei (AGN) are 
intriguing cosmic phenomena which remain subject of intensive observational 
and theoretical study since the day of their discovery. In spite of the tremendous 
progress both in observations and theory, we are still some way from solid  
understanding of their physics.     
One of the most remarkable properties of cosmic jets is their ability to 
keep structural integrity over huge distances. Consider for example jets from 
young stars. They are traced up to distances of few parsecs and their 
initial radius, which should be about the size of their central engine, 
is somewhere between the stellar radius of $\simeq 2 R_\odot$ and $0.1-10\,$AU,
depending on the engine model \citep{ray-12}. Thus, stellar jets cover the distances 
of order $10^5$ or $10^7$ of initial radii. 
The data for AGN jets are even more impressive. Assuming that they are powered 
by the Blandford-Znajek mechanism \citep{BZ-77}, the initial jet radius is 
comparable to the gravitational radius of the black hole. For the typical 
value of the black hole mass, $10^9 M_\odot$, this is $r_i\sim 10^{14}\,$cm.        
These jets can be traced up to the distances of hundreds kilo-parsecs, which is 
about one billion(!) initial radii. None of the jets produced in laboratories
using most sophisticated jet engines come even close to their cosmic counterparts
in terms of their ``survival'' abilities. They loose integrity and get destroyed by 
dynamic instabilities on much smaller scales, no more than a hundred of initial jet radii.    

This remarkable apparent stability of cosmic jets has attracted a lot
of attention from theorists, resulting in a very long list of
analytical and numerical studies.  A comprehensive review of these studies is 
beyond the scope of this introduction, for this we refer the interested reader 
to the recent reviews by \citet{hardee-11} and \citet{perucho-12}.
Here we only outline some key concepts and results.  

Most analytical and numerical studies of jet stability are focused on flows with 
cylindrical geometry, because they are easier to analyze. 
The main instabilities in such flows are \textit{1.} \textit{Kelvin-Helmholtz
instability} (KH) where the source of energy for unstable modes is the bulk motion 
of the flow \citep[e.g.][]{Birkinshaw1991} and \textit{2.}
\textit{Magnetic instabilities} which utilize the energy of the jet magnetic field.   
The latter are similar to those  encountered in the experiments on magnetic
confinement of plasma \citep{Bateman:1978}. The magnetic instabilities are 
important as most astrophysical jets are believed to be produced via a magnetic 
mechanism.  

In a static column, ideal\footnote{
Due to the enormous Reynolds numbers of astrophysical jets, only ideal effects need 
be considered for the overall stability of the flow.} 
MHD instabilities are often
classified as either \textit{current driven} (CD) where the quantity 
$\mathbf{j\cdot B_0}$ determines the outcome, or \textit{pressure
driven} (PD) where $\nabla p_o \approx \mathbf{j \times B_0}$ is important  
\cite[see e.g.][]{Freidberg1982}. 
In a magnetised flow, the separation between the KH and magnetic instabilities 
is not that rigid as they can mix and give  rise to a new phenomenon 
\citep[e.g.][]{BatyKeppens2002}.  

Whatever the nature of the instability is, the most disruptive mode is the kink mode ($|m|=1$), 
which leads to a displacement of the center of mass in the jet cross section. It is recognized 
that the kink mode can come in two forms -- the internal one, where the jet boundary is fixed 
(e.g. due to a rigid wall) and the external one where it is perturbed as well. Only the latter 
one is a danger for the jet integrity. In the astrophysical context, the internal mode can 
actually be beneficial, leading to dissipation required by observations.

Linear stability analysis of cylindrical MHD jets show that they are generically 
unstable to the kink mode. Various factors have been shown to influence 
the growth rate: jet Mach number, density of external medium, velocity shear in 
the jet, magnetic structure, relativistic effects etc, but none seem to lead to full 
stabilization under conditions appropriate for astrophysical jets 
\citep{2007ApJ...662..835M,2002ApJ...576..204H,2003ApJ...583..116H, 
2002ApJ...576..204H}. 
However, the mere fact that a jet is linearly unstable does not necessary mean that 
it will be completely destroyed by the instability. Its growth may  
saturate at non-linear phase rather early and result only in mild deformations.
Numerical simulations are normally required to handle the non-linear phase and give 
insight on the saturation regime. In most numerical studies so far, the kink instability
shows to be highly disruptive.  However, for force-free magnetic configurations  
the non-linear growth rate can be relatively low \citep[e.g.][]{OBB-12}. 
These configurations require poloidal magnetic field comparable to the azimuthal one, 
which is not feasible at large distances from the jet source. Near the source the 
poloidal magnetic field may provide the required stability, as indicated by the 
numerical simulations \citep{BM-09,Porth2013}.

Outflows from neutron stars and black holes can be highly magnetically dominated,
with magnetic energy density significantly exceeding that of the rest-mass energy 
of plasma. In the limit of zero inertia of plasma, the equations of Relativistic MHD 
reduce to that of Magnetodynamics (MD), where plasma influences the dynamics of 
electromagnetic field only via perfect conductivity \citep{ssk-ffde}. 
Analytical studies of MD cylindrical jets concluded that under some conditions they 
can be stable. In particular, \citet{istomin1994} demonstrated stability to 
internal kink mode in a jet with uniform axial magnetic field ($B_z=$const), 
and \citet{Lyubarskii:1999} showed that an unbounded flow is stable if $B_z$ 
does not decrease outwards. \citet{narayan2009} considered quasi-cylindrical 
equilibrium with an extra term accounting for finite curvature of magnetic 
surfaces and concluded that it is stable to the internal kink mode provided 
the flow speed (the drift speed in MD) increases outwards. $B_z$ is uniform in 
their equilibrium. The applicability of the MD approximation is rather limited.
The inertial effects become important when the flow becomes super-fast 
magnetosonic. This occurs when the flow Lorentz factor exceeds 
$\Gamma_f \approx \Gamma\sb{max}^{1/3}$, where $\Gamma\sb{max}$ is the terminal 
Lorentz factor corresponding to complete conversion of magnetic energy into 
the bulk motion energy. Thus MD is justified only for the small initial section 
of the acceleration zone.

The observations of jets from young stars, X-ray binaries, micro-quasars, gamma-ray 
bursts and active galactic nuclei tell us that their physical parameters differ enormously. 
One could use this to argue that there exists no single cause for the stability of the 
various types of jets. However, it would be much more satisfying to have a universal 
mechanism. In this case, the explanation must be very robust and simple and it must be 
build around one property common to all these flows. 
One such common property is the rapid lateral expansion of cosmic jets. With the opening 
angle of few degrees, the jets of young stellar objects must expand laterally by a factor 
exceeding $10^4$. The radius of the M87 jet near its tip is $r_j\sim\mbox{few}\times 10^{20}$cm, 
indicating the total increase of the jet radius by about $10^6$(!) times. Such a dramatic 
expansion stems from the fact that cosmic jets originate from compact objects, 
whose gravitational field induces rapid decline of pressure in their surrounding. 
This is a natural reaction of collimated super-sonic 
flows to the drop of external pressure in their attempt to establish transverse 
force equilibrium. It has already been pointed out that expansion has a 
stabilizing effect on jets dynamics \citep[e.g.][]{RosenHardee2000,MSO-08}. 
In our paper, we argue that this is in fact the main reason behind the apparent 
enhanced stability of cosmic jets.

The issue of stability of cosmic jets in not limited to the problem of their survival. 
It has been long recognized that emissivity of adiabatically expanding flows drops much 
faster compared to what is observed in cosmic jets \citep{BBR-84} and some ``in-situ'' 
dissipation and particle acceleration is required to explain the observations 
\citep[e.g.][]{FerrariTrussoni1979,sikora2005,Meisenheimer2003,BrunettiMack2003}. 
Internal shocks caused by variability of the central engine and interaction with 
the environment have been often invoked to introduce such dissipation. However, 
this may not be sufficient. This is particularly problematic for relativistic
shocks in magnetised plasma as the recent Particle-in-Cell simulations revealed that 
they are not efficient non-thermal particle accelerators \citep{ss-09,ss-11}. A viable 
alternative to shocks is the magnetic dissipation associated with magnetic 
reconnection \citep[e.g.][]{SDD-01,SS-14,SPG-15}. 
This however requires development of thin current 
sheets, which may occur naturally via instabilities. Instabilities may also lead to 
formation of shocks. Such instabilities must be strong and yet not threatening to 
the jet integrity. These seemingly conflicting requirements can only be met by 
local internal instabilities, developing on a small scale compared to the jet radius and 
arising from finer structures inside the jet.  In this paper, we give an example of 
such instability occurring in the magnetically confined jet core.

The observations suggest that global instabilities may also play a role. For example, some 
cosmic jets appear quite ``wiggly'' \citep[e.g.][]{CB-96}, implying an 
external kink mode at work. 
Moreover, the observed properties of Fanaroff-Riley type I (FR-I) extragalactic radio sources 
indicated that their jets become completely destroyed by instabilities, mix with the 
external gas and turn into buoyant turbulent plumes \citep[e.g.][]{bicknell-84}. 
We argue that this loss of 
jet global stability can occur when it enters regions where the external pressure 
distribution flattens out.  In the case of AGN jets this can be the core of the X-ray 
corona of the parent galaxy or the extended radio lobe.

This paper is organized in the following way. In Section~\ref{sec:SaC}, we put forward
very simple and general arguments, which explain how lateral expansion increases 
global stability of jets. In brief, such expansion slows down, and may even completely 
terminate, the flow of information across jets, thus reducing the growth of coherent 
displacements. The rate of expansion depends on the properties of jet surrounding, namely 
on how fast the external pressure decreases with distance from the jet origin. For power-law 
atmospheres, $p_{ext} \propto z^{-\kappa}$, there is a critical value for the power index, 
$\kappa=2$. For steeper gradients, causal connectivity is lost and the jets are globally stable. 
In astrophysical context, such steep atmospheres are expected to be quite common.         
The generality of the argument makes this a very robust and hence attractive explanation, 
but details depend on the actual internal jet structure. Numerical simulations 
are required to study the non-linear development of instabilities, particularly in the 
sub-critical regime with $\kappa<2$. Our efforts in this direction are   
described in Section~\ref{sec:pbs}, where we focus on a particular 
class of magnetised relativistic jets, whose initial internal structure is described by
the core-envelope model of cylindrical jets due to \citet{ssk-mj99}. 
These jets have a z-pinched inner core and 
a force-free envelope with purely azimuthal magnetic field. 
A simple method to obtain initial near stationary solutions of relativistic expanding jets was presented in \cite[][thereafter KPL]{KPL2015}.  
In Section~\ref{sec:pbs},  
this approach is generalized in a way which allows us to study the time-dependent 3D dynamics 
of these flows using periodic box simulations. In both cases, the jet expansion is triggered by 
a gradual lowering of the external gas pressure, which imitates the conditions experienced by 
the jet material as it propagates through power-law atmospheres. 
In Section~\ref{sec:discuss}, we discuss the astrophysical implications 
of our findings. In particular, we propose that the division of extragalactic radio source 
into FR-I and FR-II classes is related to the stability issue. The lower power FR-I jets are 
externally confined in the coronas of their host galaxies, which have rather flat pressure 
distribution, do not expand sufficiently rapidly, become unstable and mix with the coronal 
plasma on the galactic scale. In contrast, the more powerful FR-II jets remain free 
and stable until they reach the scales of radio lobes. The instability of jet 
cores may cause their disintegration and trigger internal dissipation and ultimately 
electromagnetic emission even when the envelope is stable. This may explain the emission 
of FR-II jets on much smaller scales than radio lobes. Our conclusions are summarised in 
Section~\ref{sec:concl}.

\section{Stability and Causality} 
\label{sec:SaC}
      
During the development of instabilities that may threaten the jet integrity, 
global perturbation modes are amplified. These modes involve coordinated motion 
of the whole jet and hence imply communication between its different parts by 
means of waves. These waves trigger forces that push the flow away from its equilibrium 
state. In the case of unmagnetised fluid, these are sound waves. In magnetised fluids,  
these are mainly fast magnetosonic waves. If the whole section of the jet is to be 
displaced to one side of the original jet axis, as this occurs in the kink mode, 
any part of this section needs to ``know'' what the other parts do. In other words, 
the jet has to be causally-connected in the direction transverse to its direction 
of motion. In the case of super-sonic (or super-fast-magnetosonic) flow, all these 
waves are advected with the flow and the region of influence of any particular point 
has the geometry of a cone, aligned with the flow direction. In fluid dynamics, this 
cone of influence is known as the {\it Mach cone}.  For such flows, the causal 
communication in the transverse direction is obstructed -- no wave can originate 
at one edge of the cross section and reach the other. 
However, a synchronised motion may still be possible
as long as an upstream location can be communicated with the whole of 
the jet somewhere downstream. The higher the Mach number, the longer the 
separation along the jet between its causally connected sections and slower 
the growth of unstable perturbation modes become.  This is the reason why supersonic flows 
are less unstable. 

For cylindrical jets, this necessary condition for the global instability is always 
satisfied. This explains why laboratory and terrestrial jets are relatively 
quickly destroyed by instabilities and why the theoretical studies of flows with 
cylindrical geometry struggle to explain the stability of cosmic jets. 
For expanding flows, the situation is more complicated as now there is a 
competition between the jet expansion and the expansion of the cone of 
influence. Let us analyse this competition in the simple case of a power-law 
atmosphere, with pressure $p_{ext}\propto z^{-\kappa}$.  

For a start, consider an unmagnetised non-relativistic highly supersonic adiabatic jet. 
Denote as $\theta_j=r_j/z$ its half-opening angle and as $\theta_M$ its Mach angle.         
In the limit of small angles, $\theta_M=a/v$, where $a$ and $v$ are the sound and 
bulk motion speeds of the jet respectively. In such a flow, $v=$const, 
$\rho\propto r_j^{-2}$ and $p\propto\rho^{\gamma}$, which leads to 
${\theta_M}/{\theta_{j}} \propto z \sqrt{p}$. Finally, using $p=p_{ext}$,  one 
finds that      
\beq               
  \frac{\theta_M}{\theta_{j}} \propto z^{(2-\kappa)/2}\, .  
\label{eq1}    
\eeq        

Magnetic field introduces an additional degree of complexity as the magnetic 
hoop stress can result is strong axial pinching of the jet and hence a mismatch 
between the internal jet pressure and the external one. This is particularly true for 
magnetically-dominated jets, where the magnetic pressure dominates over the thermal one.
For scale-free external pressure, one would expect the jets to be self-similar 
and hence $B^2 \propto p_{ext}$. Away from the central engine the magnetic field is 
mainly azimuthal and evolves as $B\propto r_j^{-1}$, whereas $\rho\propto r_j^{-2}$ 
as before. Thus, the Alfv\'en speed $c_a^2\propto B^2/\rho \propto r_j^0 $,  
the Mach angle based on the Alfv\'en speed $\theta_M\propto r_j^0$ and the opening angle
$\theta_j^{-1} \propto z \sqrt{p_{ext}}$.  The last two results ensure that 
equation~(\ref{eq1}) still holds in this limit. 

The relativistic case is a little bit more complicated as even in the hyper-sonic regime 
the thermal energy may dominate the rest mass energy of gas particles, $p\gg\rho c^2$,  
and the jet may continue to accelerate.  Combining the energy conservation, 
$ p\Gamma^2 r_j^2 = \mbox{const}$, and the mass conservation 
$\rho\Gamma r_j^2 =\mbox{const}$, where $\Gamma$ is the jet Lorentz factor and 
$\rho$ is its co-moving density,  with the equation of state $p\propto\rho^\gamma$, one 
finds $\Gamma \propto p^{(1-\gamma)/\gamma}$, $r_j\propto p^{(\gamma-2)/2\gamma}$, whereas 
the sound speed is constant, $a=c\sqrt{\gamma-1}$, and the relativistic Mach angle 
$\theta_M\propto 1/\Gamma$. Taken together, these yield equation~(\ref{eq1}) again.  
Thus, Eq.(\ref{eq1}) is quite general, 

The form of Eq.(\ref{eq1})
suggests that $\kappa=2$ is a critical value -- for $\kappa<2$ the jet can remain 
causally connected, whereas for $\kappa>2$ the connectivity will be lost.  
In order to verify this conclusion, we consider a flow characteristic that originates
at the jet boundary and moves towards its axis. Its equation is  
\beq 
   \oder{r}{z} = \theta_v-\theta_M\, ,
\label{eq2}
\eeq  
where $\theta_v=(r/r_j)dr_j/dz$ is the local streamline angle. Given the lack 
of characteristic length scale, one may assume that all the jet parameters are 
powers of $z$. In particular, $r_j\propto z^\alpha$ with  $0<\alpha<1$ and 
$\theta_M \propto z^\beta$. Given these, the solution of Eq. (\ref{eq2}) is 
\beq
  r = A z^\alpha \left(1-\frac{C}{\delta} z^{\delta}\right) \, ,
\eeq
where $A,C>0$ are constant and $\delta=1+\beta-\alpha$. 
One can see that the characteristic eventually reaches the 
jet axis only when $\delta>0$ -- this is a new form of the connectivity condition.
In order to turn this condition into the condition of the  power index 
of the atmosphere, we note that 
\beq
    \frac{\theta_M}{\theta_j}\propto z^{\delta}\,.  
\eeq
Comparing this with eq.(\ref{eq1}), we identify $\delta=(2-\kappa)/2$ and hence 
confirm that $\kappa=2$ is indeed a critical value. 
 
The case of relativistic Poynting-dominated flows is more complicated as the flow is 
generally not self-similar. In fact, a gradual redistribution of poloidal 
magnetic flux across the jet is an essential 
component of the so-called collimation-acceleration mechanism
\citep[e.g.][]{kvkb-09,ssk-marj}. However, the analysis presented in \citet{kvkb-09} shows 
that $\kappa=2$ is still a critical value. 

When $\kappa>2$, jets become free, with conical geometry of streamlines. 
Their pressure decreases 
rapidly -- at least as $z^{-2\gamma}$ in the gas pressure-dominated regime and 
$z^{-4}$ in the magnetic pressure-dominated one. When this is faster than the external 
pressure, a reconfinement shock can be driven inside the jet \citep[e.g.][]{sanders-83,KF-97}. 
Since shock waves are faster 
compared to sound waves, one may wonder if they can establish pressure balance with 
the external gas via dissipative heating of jet plasma and support jet connectivity. 
The key question is whether the reconfinement shock can travel all the way from the jet 
boundary to its axis. This problem has been analysed in \citet{KF-97} for the case of
unmagnetised uniform relativistic jet and it was found that the shock reaches the axis 
at the distance
\beq
   z_r \simeq \delta^{1/\delta} \fracp{L}{a\pi c}^{1/2\delta}\,,
\label{eq:rs}
\eeq       
where $\delta=(2-\kappa)/2$ and $a$ is the coefficient of the law $p_{ext}=az^{-\kappa}$ 
and $L$ is the jet power. 
It is easy to see that $z_r\to\infty$ as $\kappa\to 2$, which allows us to conclude that 
for $\kappa>2$ the jet still remains causally disconnected. Although this analysis is restricted 
to a particular type of flow, the conclusion must be generic\footnote{The dependence of $z_r$ on 
$L$, $a$, and $c$ can be recovered from the analysis of dimensions. For non-relativistic jets, 
$c$ has to be replaced with the jet speed.}.  Indeed, the dynamic pressure 
of any free jet decreases as $\propto z^{-2}$ and always wins the competition with the 
external pressure when $\kappa>2$.      

\begin{figure*}
\includegraphics[height=8cm]{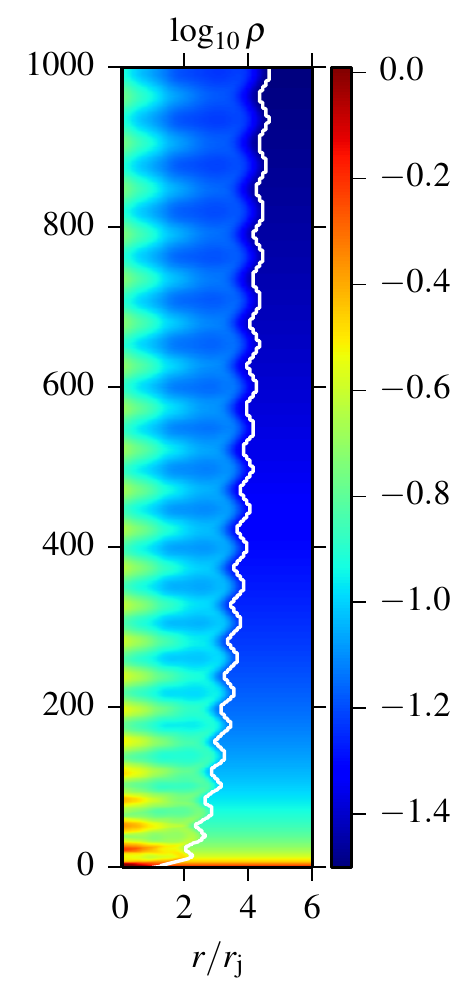}
\includegraphics[height=8cm]{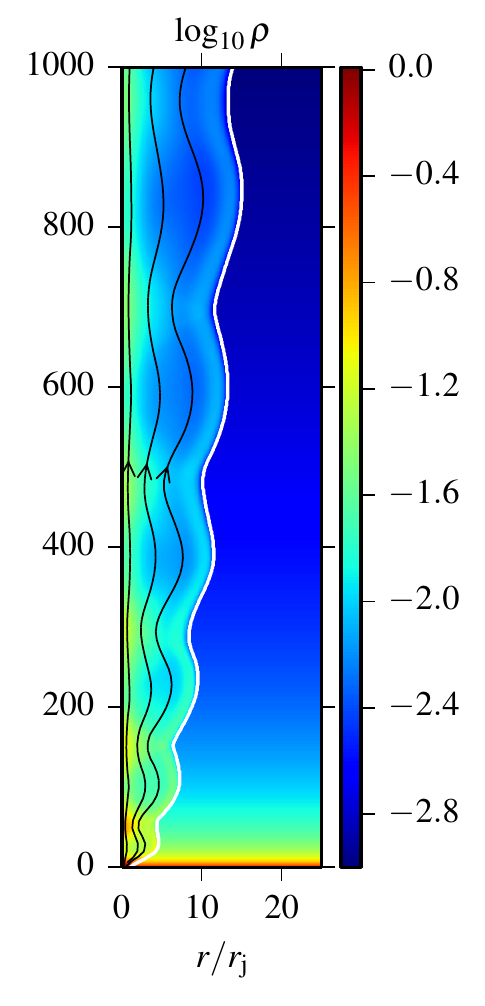}
\includegraphics[height=8cm]{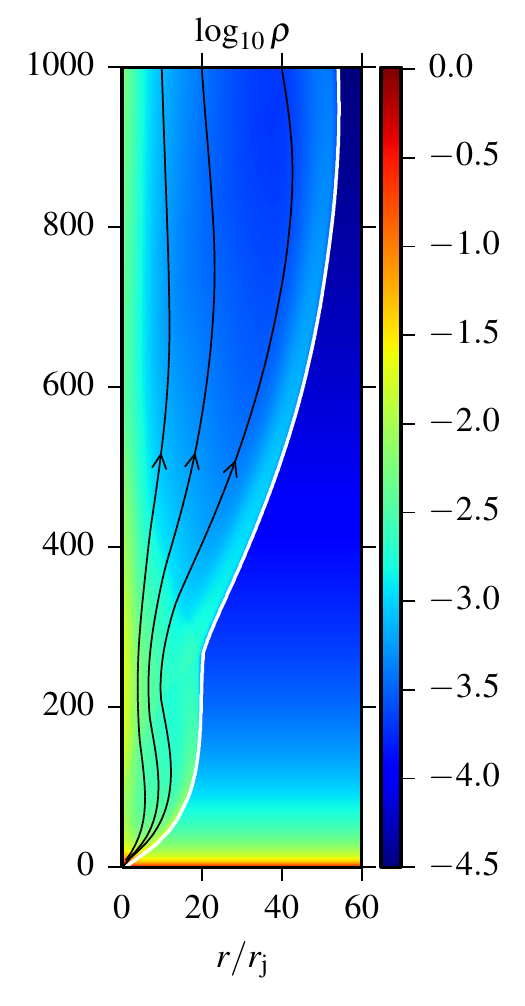}
\includegraphics[height=8cm]{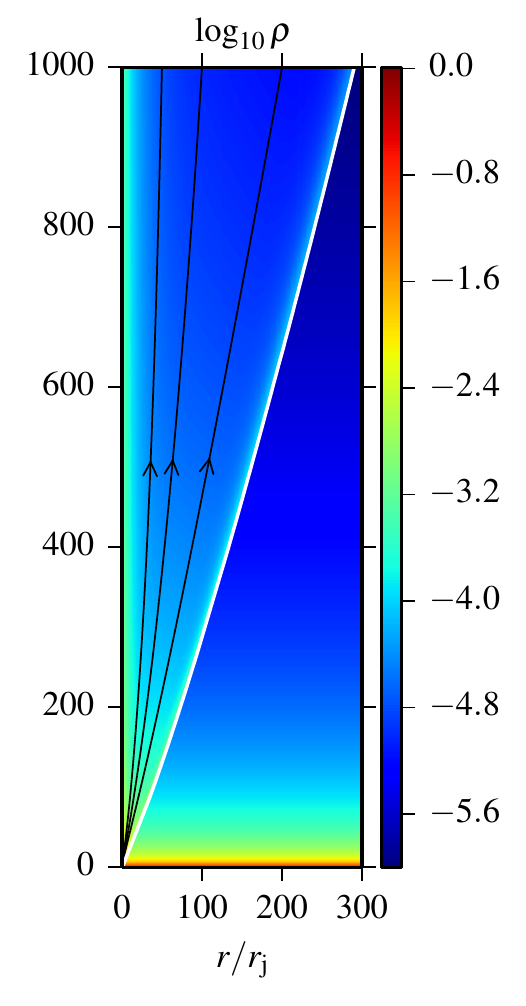}
\includegraphics[height=6cm]{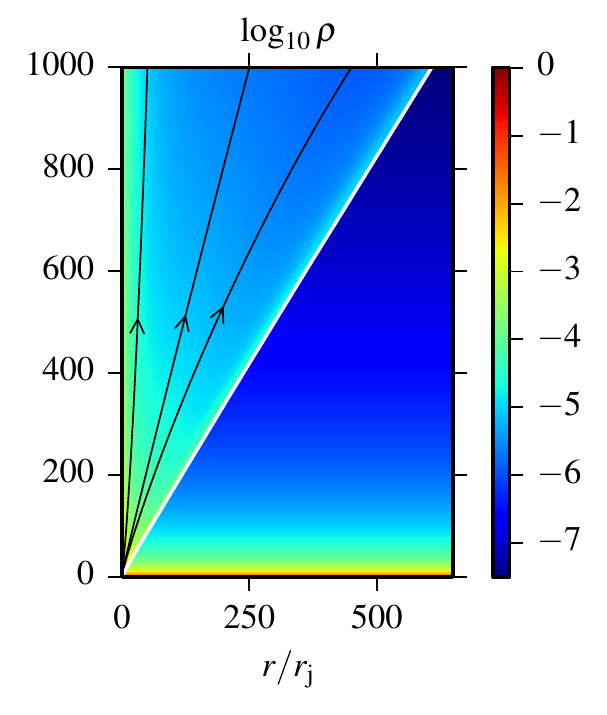}
\caption{Structure of steady-state jets obtained via time-dependent 1D simulations.
The plots show the density distribution for models with $\kappa=0.5$, 1.0, 1.5, 2.0 and
2.5, increasing from left to right. The distance along the vertical axis is defined as
$z=ct/r_j$, where $r_j$ is the initial jet radius. The white contour shows the jet boundary,
located using the passive scalar. }
\label{fig:1d}
\end{figure*}

To confirm and illustrate this analysis, we constructed a set of steady-state jet solutions
for jet propagating in a power-law atmosphere using the method described in KPL. 
The jet structure at the nozzle (of radius $r_0=1$) is described in Sec.\ref{sec:initial} below. 
It represents an equilibrium cylindrical flow in pressure balance with the external medium. 
Figure~\ref{fig:1d} shows the results for models with
$\kappa$ ranging between 0.5 and 2.5. First, one can see how the overall jet shape 
changes from an almost cylindrical for $\kappa=0.5$ to a conical for $\kappa=2.5$. 
Second, these plots nicely illustrate how the jets are trying to maintain the
transverse equilibrium by means of magnetosonic waves bouncing across the jet. 
These waves are launched due to the loss of dynamic equilibrium downstream of the 
nozzle, where the jet interior becomes under-expanded because of the drop in the 
external pressure, and cause the oscillations of the jet boundary about the mean position.  
As $\kappa$ increases, the wavelength of the oscillations increases as well until 
they disappear for $\kappa\ge 2$. At this point the causal connectivity across the jet 
is completely lost.

How does this critical value compare with those of typical environments of cosmic jets?
For a polytropic atmosphere of central mass, one has $\kappa=\gamma/(\gamma-1)$,
which is higher than 2 when $1<\gamma<2$. 
For a spherical adiabatic wind, $\kappa=2\gamma$, which is also 
steeper than the critical one ( 
Only a self-collimating magnetic wind can in principle deliver $\kappa<2$.). 
For the Bondi accretion $\kappa =3\gamma/2$, which is still larger than 2 for 
$\gamma>4/3$. Thus, steep gradients of external pressure, bordering the critical 
value, are expected to be quite common close to the central engine. 

Observational measurement of gas and magnetic pressure in 
the environment of cosmic jets is not yet always possible, particularly close to the source. 
Most of the time, only indirect model dependent estimates can be made.  
Taken together, all the estimates available for 
AGN show that $\kappa\simeq2$ is a typical mean value for their environment 
\citep{phinney-83,BBR-84}. To illustrate the arguments, consider the conditions at 
the jet source. For quasars, we are dealing with $M\simeq 10^9M_\odot$ black holes, 
which accrete at the rate close to the Eddington's one. The inner parts of their 
accretion discs are thus dominated by radiation pressure and have constant thickness 
\begin{equation}
    H \simeq \frac{3}{2} \fracp{\dot{M}c^2}{L_{edd}} r_g,
\end{equation}
where $L_{edd}\simeq 1.3\times 10^{47}(M/10^9M_\odot)\,$erg/s 
is the Eddington luminosity of mass $M$ and $r_g=GM/c^2$ is its 
gravitational radius \citep{SS-73}. Thus near the black hole, the disk becomes    
geometrically thick and its vertical hydrostatic equilibrium yields 
the pressure estimate $p=GM\rho/H$. The gas density can be found from the mass 
conservation as $\rho=\dot{M}/2\pi r H v_{\rm r}$, where in the $\alpha$-disk model 
the accretion speed $v_{\rm r} = \alpha a_{\rm s} H/r$ and
$a_{\rm s}^2\simeq p/\rho$. Combining these equations, we find
\begin{equation}
 p= \frac{\dot{M} (GM)^{1/2}}{2\pi\alpha H^{5/2}}. 
\label{eqn:disk-p}
\end{equation}
For a $10^9M_\odot$ black hole, this gives us the gas pressure 
$p_{disk}\simeq (10^7/\alpha)$dyn/cm$^2$ at the scale of order 
$r_g \simeq 1.5\times 10^{14}$cm. On the other extreme, 
inside the extended radio lobes of the size $\simeq 100\,$kpc, 
the typical pressure inferred from the radio and X-ray observations is of the order 
$p_{lobe}\simeq 10^{-11}$dyn/cm$^2$. Assuming these are the end points of a 
single power-law, we find its index $\kappa\simeq 2$.  

Thus, both the theoretical and observational arguments indicate that rapid  
decline of pressure with distance from their source must be typical for cosmic jets
and their corresponding lateral expansion should be fast enough to
make a strong positive impact on their global stability.         
Given the huge range of scales, it would be unrealistic to expect the same slope 
everywhere. In fact, in hot coronas of elliptical galaxies $\kappa = 1.25\pm.25$ 
(Mathews \& Brighenti 2003). 
Moreover, inside the extended radio lobes, which expand much slower compared to their 
sound speed, one would expect $\kappa\simeq 0$. Within such 
flat sections global instabilities may develop, leading to the observed flaring and 
wiggling of cosmic jets, whereas through steeper sections they pass mainly undisturbed.

\section{Periodic box simulations of expanding jets}
\label{sec:pbs}

To study the jet stability, one has to carry out fully
three-dimensional simulations, as the most threatening mode is a
non-axisymmetric $m=1$ current driven mode
\citep{begelman1998,appl2000}. For supersonic
jets, such instabilities are waves travelling in the general direction of the jet flow 
and experiencing growth in amplitude downstream. The best way of studying their development 
is to use very long computational domain, exceeding in size the initial jet radius by 
several orders of magnitude. Since such simulations are computationally very expensive, 
much smaller computational domains which capture only a section of the jet have been used 
in many computational studies instead 
\citep[for some of the recent examples see ][]{MLNH-12,AMBR-14}.
To allow travelling waves, such domains are combined with periodic boundary 
conditions at the boundaries normal to the jet axis.  
In such a ``periodic box'', all waves that leave the computational domain through one of 
the periodic boundaries, enter it through the opposite one. 
Obviously, such simulations allow to study only modes whose wavelength 
is below the box size. To be more precise, a multiple of the wavelength must be equal 
to the box length. The box frame does not have to be stationary relative to the jet 
source - it may well be moving relative to it. Clearly, the periodic box simulations 
are best suited for studying the instability of cylindrical flows ($\kappa=0$). 
In order to study the role of the jet expansion in atmospheres with 
$\kappa>0$, one may force the external pressure in the box to decrease in a systematic 
fashion, thus triggering the jet expansion in the radial direction. This is exactly the 
approach we apply in our simulations.

\subsection{Initial conditions}
\label{sec:initial}

As starting point of our investigation, we choose a cylindrical plasma column in equilibrium between Lorentz forces and pressure gradient.  The velocity is directed in vertical direction $v_r=v_\phi=0$ and the field is purely toroidal. Thus we have the simple relation 
\beq
  \oder{p}{r} + \frac{b^\phi}{r} \oder{rb^\phi}{r} =0\,,
\label{eq-eqv}     
\eeq
where $b^\phi=B^\phi/\Gamma$ is the azimuthal component of the magnetic field as measured 
in the fluid frame using a normalized basis.  
One of the infinitely many solutions of (\ref{eq-eqv}) is the ``core-envelope'' model of \citet{ssk-mj99}: 
\beq
b^\phi(r) = \left\{
\begin{array}{ccl}
  b_m(r/r_m) &;& r<r_m \\ 
  b_m(r_m/r) &;& r_m<r<r_j
  \\ 0&;& r>r_j
\end{array}
\right. , 
\label{eq:bphi}
\eeq
\beq
p(r) = \left\{
\begin{array}{ccl}
  p_0\left[\alpha+\frac{2}{\beta_m}(1-(r/r_m)^{2})\right] &;& r<r_m \\ 
  \alpha p_0 &;& r_m<r<r_j \\ 
  p_0 &;& r>r_j
\end{array}
\right. ,
\label{eq:pressure}
\eeq
where 
\beq
\beta_m= \frac{2 p_0}{b_m^2},\qquad \alpha=1-(1/\beta_m)(r_m/r_j)^2\,, 
\eeq
$r_j$ is the jet radius and $r_m$ is the radius of its core. As one can see, the core
is pinched and in the envelope the magnetic field is force-free. This may be combined 
with any distribution of density and axial velocity. We imposed $\rho=\rho_0$ and 
\beq
\Gamma(r) = \Gamma_0 \left(1-(r/r_j)^\nu\right) + (r/r_j)^\nu\,.
\eeq
The parameters for the initial equilibrium are $r_j=1$, $r_m=0.37$,
$b_m=1$, $\rho_0=1$, $z_0=1$, $\beta_m=0.34$, $\Gamma_0=3$ and $\nu=2$.  
The jet is only moderately magnetised with $\sigma_{\rm max}=0.7$.  

We also investigate one force-free jet model where the pressure is set
to $p=\alpha p_0$ everywhere in the jet and an additional poloidal
field
\beq
B^z =
\left\{
\begin{array}{ccl}
 p_0\left[\frac{2}{\beta_m}(1-(r/r_m)^{2})\right] &;& r<r_m\\
 0 &;& r>r_m\,.
\end{array}
\right.
\eeq
yields the pressure balance in the core.  The equation of state used
throughout this work is $w=\rho +\gamma/(\gamma-1)p$ with constant
adiabatic index $\gamma=4/3$.

\subsection{Numerical treatment}
\label{sec:numerics}

\begin{figure*}
\begin{center}
\includegraphics[width=8cm]{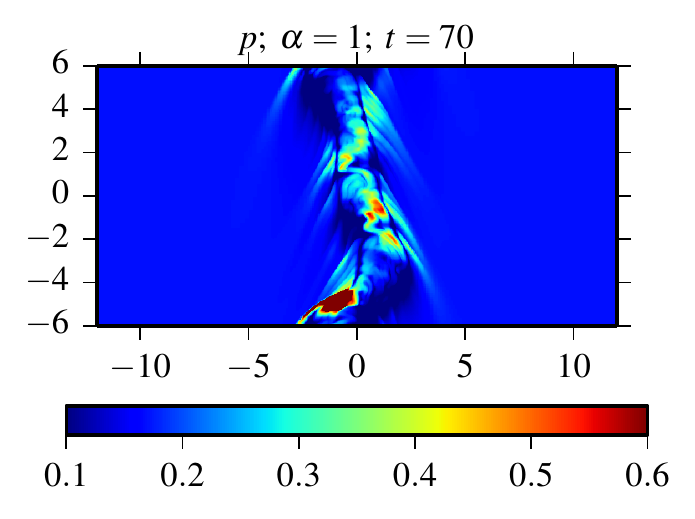}
\includegraphics[width=8cm]{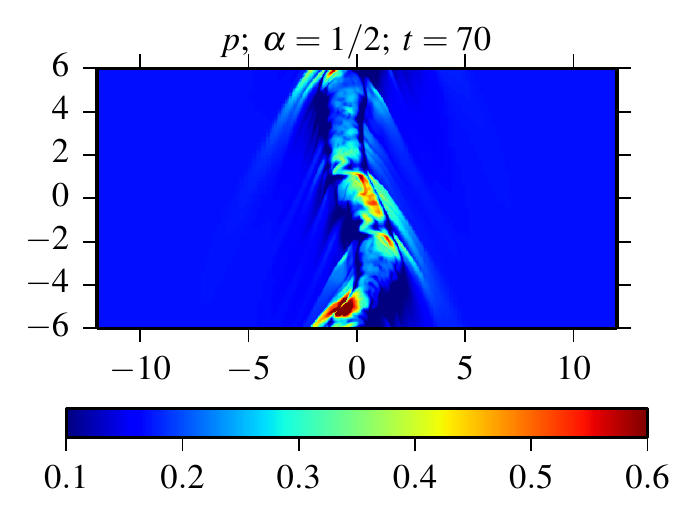}
\includegraphics[width=8cm]{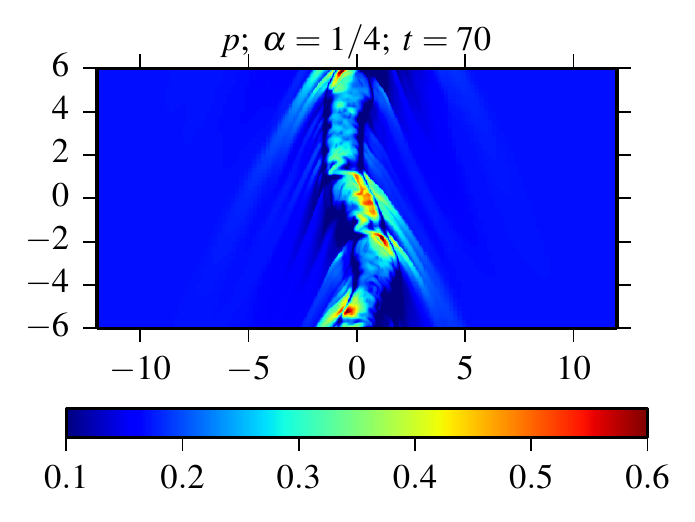}
\includegraphics[width=8cm]{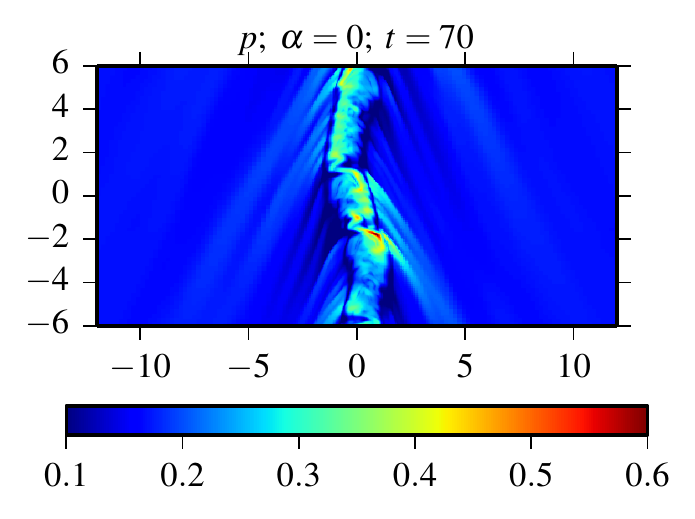}
\caption{Dependence on the dampening parameter $\alpha$.  
The plots show the gas pressure distribution in the $y=0$ plane for models 
with  $\kappa=0$ and  $\alpha=1,1/2,1/4$ and 0 at the same instance $t=70$.  
} 
\label{fig:varalpha_p}
\end{center}
\end{figure*}

So far periodic box simulations have been used only to study flows with cylindrical geometry, thus
excluding the effects of jet expansion. At first glance, this seems to be the only option 
as the periodic boundary conditions do not allow systematic variation of the external 
pressure in the jet direction. However, one can use a ``trick'' similar to that we have employed 
in our study of steady jets in KPL -- one may trigger the lateral expansion 
of the jet by forcing the external pressure to decrease. 
At first, we tried exactly the same approach as in the case 
of steady jets - direct resetting of the exterior solution to the prescribed state.
However, the results of our test experiments with the $\kappa=0$ model, where we could obtain 
the solution without forcing and use it as a reference, have shown that this is not quite 
satisfactory -- in the model with forcing the instability turned out to be significantly more 
violent.  The resetting amounts to complete erasing of the instability wave structure 
outside of the jet. A less drastic approach is to drive the exterior solution towards the 
desired state according to the relaxation equation
\beq
\frac{df}{dt} = -\alpha \frac{c}{r_{jet}} (f-f_{ext}(t)) \,,
\label{eq:driver}
\eeq
where $f_{ext}(t)=p_0 (ct/z_0)^{-\kappa}$ is a target value of the undisturbed state 
of external gas, $f$ is the actual current value of pressure and $\alpha (c/r_{jet})$ is the 
relaxation rate. The relaxation rate determines how far the instabilities 
can penetrate into the jet environment. The method applied in KPL is recovered in the 
limit $\alpha\to\infty$.\footnote{This forcing approach can be applied in the 1D simulations 
described in KPL. This leads to dampening of jet oscillations but 
the overall expansion rate is well preserved}

\begin{figure}
\begin{center}
\includegraphics[width=8.cm]{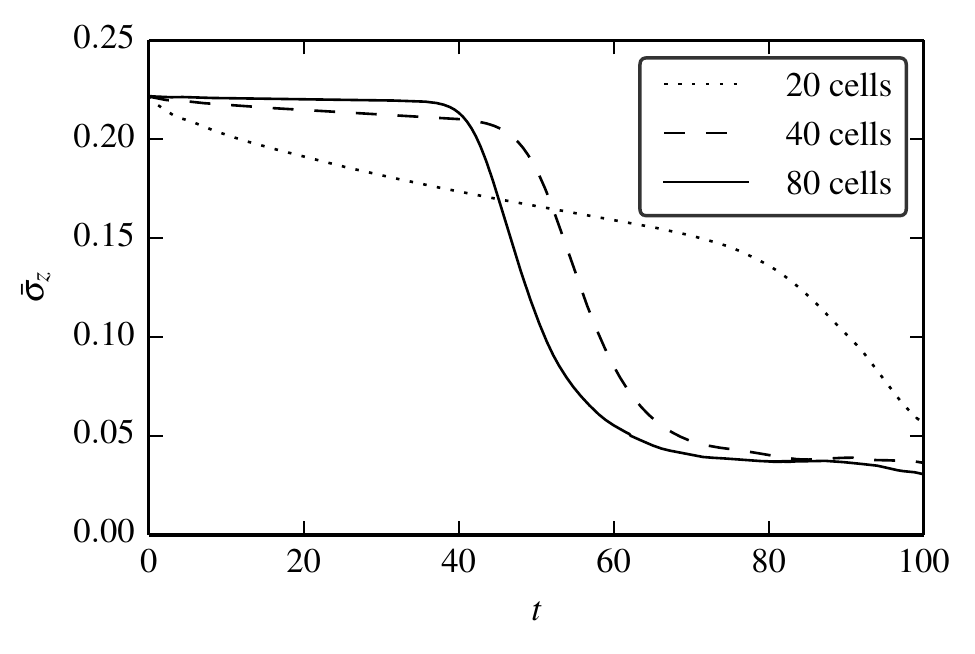}
\caption{Resolution study of the fiducial parameters with $\kappa=0$.
  Resolution is indicated in terms of cells per jet radius $r_{\rm j}=1$. }
\label{fig:resolution}
\end{center}
\end{figure}

\begin{figure*}
\begin{center}
\includegraphics[width=7.cm]{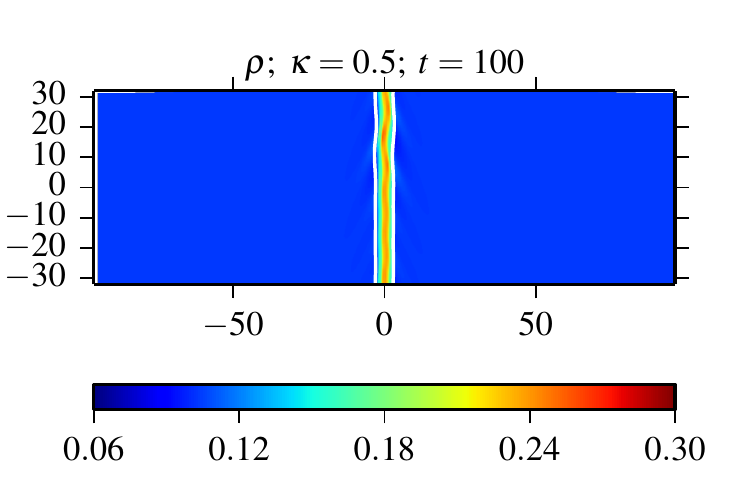}
\includegraphics[width=7.cm]{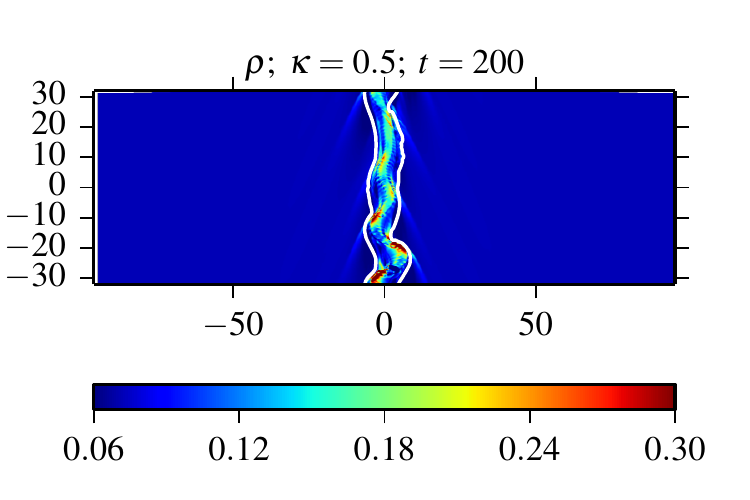}
\includegraphics[width=7.cm]{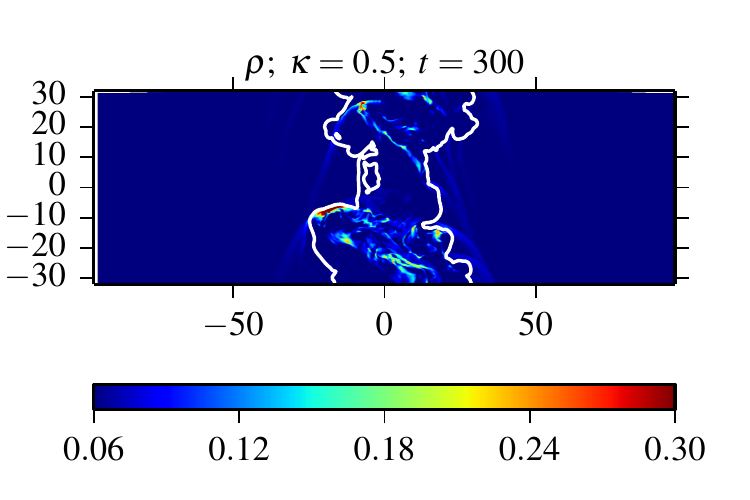}
\includegraphics[width=7.cm]{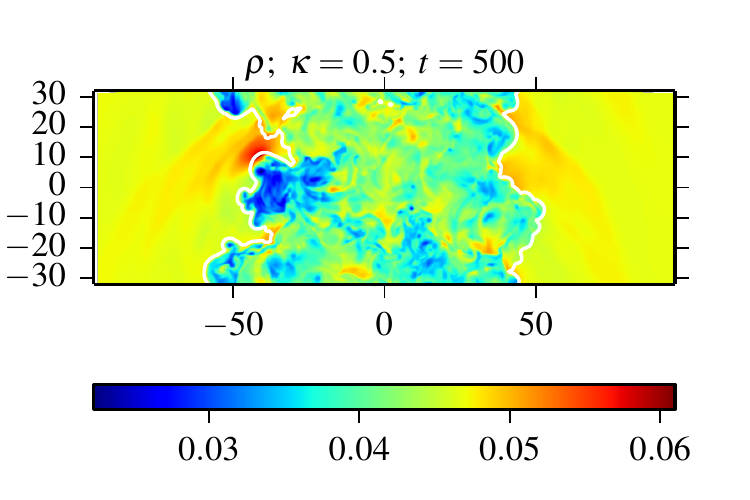}
\caption{Time evolution of density for the model $\kappa=0.5$.  Shown
are $y=0$ slices for times $t\in\{100,200,300,500\}$.  
The white contour indicates the jet boundary.  The
non-linear development of the kink instability dramatically
increases the effective jet cross-section and seeds turbulence in
the jet medium ($t=300$).  At $t=500$ the jet is disrupted entirely
and replaced by a slowly moving plume. }
\label{fig:tevol_k0-5}
\end{center}
\end{figure*}

\begin{figure*}
\begin{center}
\includegraphics[width=5.cm]{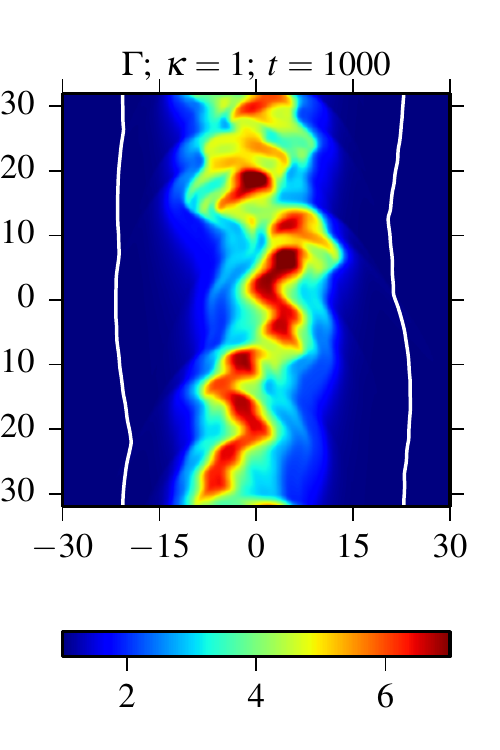}
\includegraphics[width=5.cm]{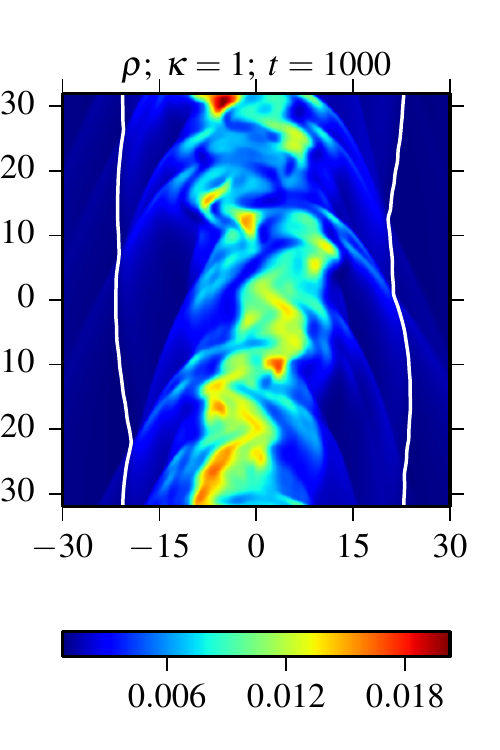}
\caption{Solution for the model with $\kappa=1.0$ at $t=1000$.  
The white contour indicates the jet boundary.}
\label{fig:3}
\end{center}
\end{figure*}

Following \citet{MLNH-12}, we perturb the initial configuration via adding the 
radial velocity component 
\beq
v^r(r,\phi,z) = \frac{\Delta v}{N} \exp(-r/r_m) \sum_{n=1}^N \cos\phi \sin(2\pi n z/L_z)\,,
\eeq
to the jet velocity field, Here $L_z$ is the box length along the jet axis and $N$ is the 
number of introduced modes.  This corresponds to an equitable superposition of modes with positive and negative azimuthal mode number $m=\pm 1$. 
In all our simulations we used $N=4$ and $\Delta v=0.01\, c$. 

In order to select the reasonable relaxation rate constant $\alpha$, we have carried 
out test simulations of cylindrical jets ($\kappa=0$) with and without 
the forcing term. Figure~\ref{fig:varalpha_p} shows the snapshots of the solutions for 
the models which differ only by the value of $\alpha$, namely $\alpha=0$, $1/4$, $1/2$ and 1. 
One can see that higher value of $\alpha$ leads to more pronounced perturbation of the 
jet structure. As a reasonable compromise, we adopted $\alpha=1/2$ for all our main runs.   

The simulations have been carried out with MPI-AMRVAC \citep{PorthXia2014,amrvac}\footnote{https://gitlab.com/mpi-amrvac/amrvac}, which 
utilizes a Godunov-type scheme with HLL approximate Riemann solver and second-order 
spatial TVD reconstruction due to \cite{koren1993}.  For the time-advance, we used a three-step 
Runge-Kutta method.  Only on one occasion, near the end of the $\kappa=2$ run, we had to 
resort to an even more diffusive Lax-Friedrich scheme to circumvent numerical issues 
related to extremely low gas pressure in the jet.  
The solenoidal condition $\mathbf{\nabla\cdot B}=0$ was  
treated by means of the \cite{Dedner02} GLM approach.

In order to determine the required numerical resolution, several 
models were run with the reference case $\kappa=0$.  
A comparison of the results for the average jet magnetization are
shown in Figure~\ref{fig:resolution}.  Runs with 40 and 80 cells per
jet radius agree quite well in both the onset of strong dissipation
in the non-linear phase at $t=40-50$ and in the saturated value
obtained at $t\simeq80$.  At 20 cells per jet radius, the solution
is markedly more diffused.  Based on this data, we have concluded
that 20 cells per jet radius is perhaps too low, whereas 80 cells
per radius is probably already an ``overkill''.  In the expanding
simulations, the jet is thus initialized with the resolution of 40
cell per jet radius on the finest AMR grid.  The maximum AMR depth
was nine levels in the $\kappa=2$ model while models with
$\kappa=0.5,\,1.0$ and 1.5 where run with seven levels.  During the
runs we refine all cells containing jet material to the current
highest level using a passive tracer $\tau$ as a jet indicator.

\begin{figure*}
\begin{center}
\includegraphics[width=8cm]{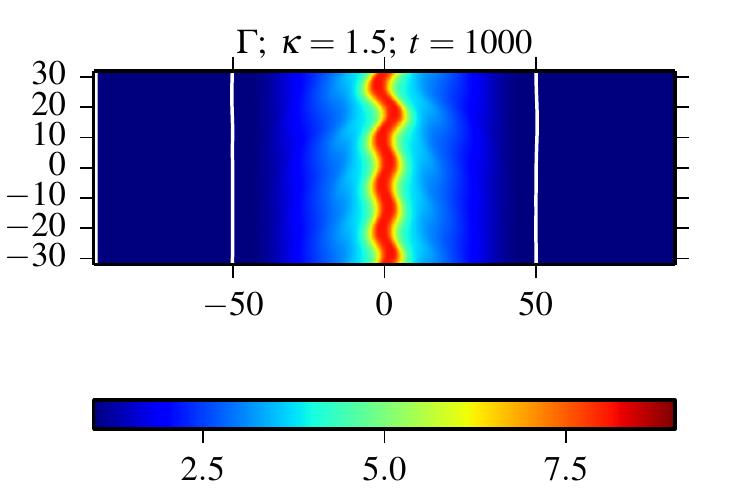}
\includegraphics[width=8cm]{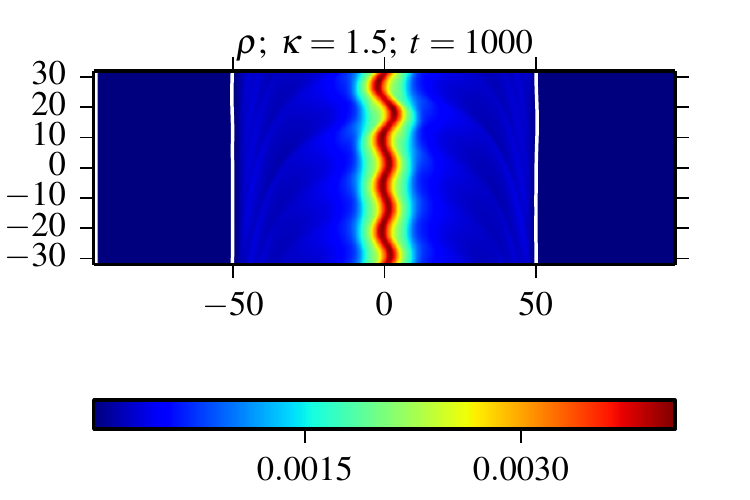}
\includegraphics[width=8cm]{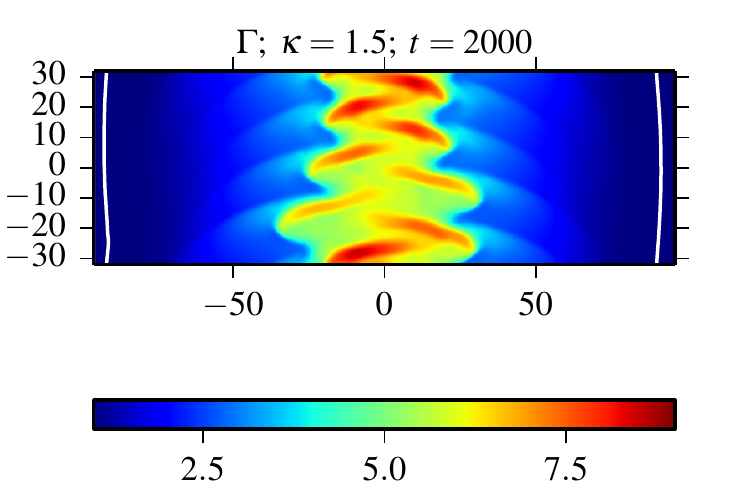}
\includegraphics[width=8cm]{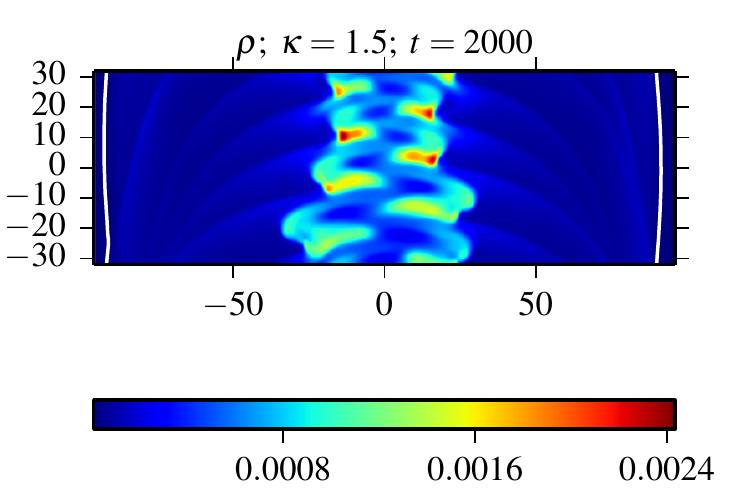}
\caption{Solution for the model with $\kappa=1.5$ at $t=1000$ (top) and at 
$t=2000$ (bottom).  The white contour indicates the jet boundary.}
\label{fig:2}
\end{center}
\end{figure*}

\begin{figure*}
\begin{center}
\includegraphics[width=8.cm]{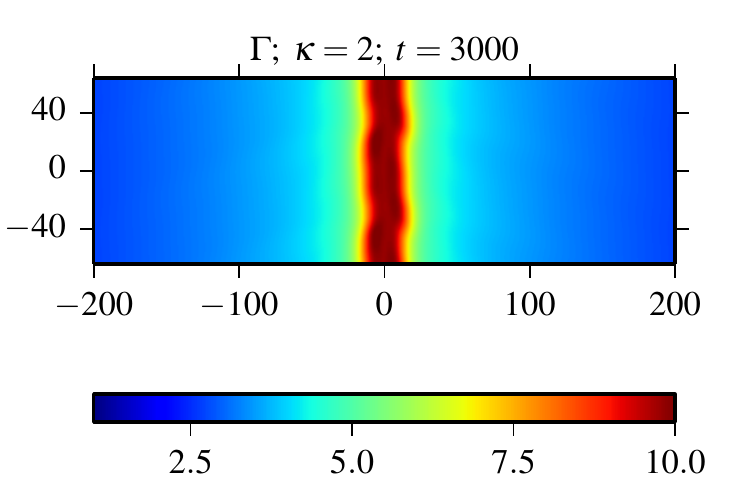}
\includegraphics[width=8.cm]{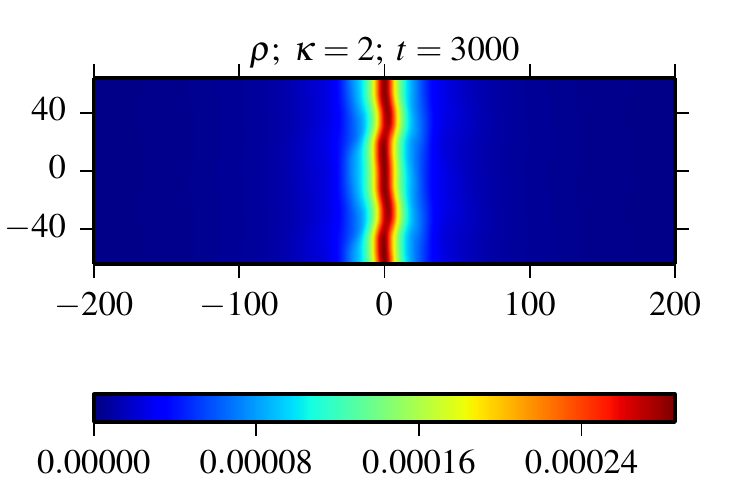}
\caption{Solution for the model with $\kappa=2.0$ at $t=3000$. The jet radius 
exceeds 200 at this point and its featureless boundary is not seen in these plots. }
\label{fig:kappa2}
\end{center}
\end{figure*}

As the jet expands, it becomes better resolved and more and more cells
need to be updated.  Thus to keep the computational cost at bay, we
coarsen the entire jet each time the jet fills more than 
$N_{\rm max}=60\times 10^6$ cells.  Assuming the jet retains its approximate
cylindrical shape, the number of cells per jet radius just after
coarsening the jet to level $l$ is
\beq \mathcal{R}=\frac{r_{\rm
    jet}}{\Delta x} = \frac{1}{2}\left(\frac{N_{\rm max} 2 \Delta x}
{\pi L_z}\right)^{1/2} \eeq
with $\Delta x = 1/40 \times 2^{l_{0}-l}$,
where $l_{0}$ is the AMR level of the jet at the start of the
simulations.  Hence after the first coarsening event, we have for
$L_z=64$: $\mathcal{R}\simeq43$, after the second 
$\mathcal{R}\simeq\sqrt{2}\times 43\simeq 61$ and so on.  Thus the resolution per jet
radius effectively increases during the course of the simulation.
At the same time, each coarsening event speeds up the simulation by a
factor of up to $16$ due to the reduced number of cells and the larger
CFL-limited time step.

The simulations were carried out in a Cartesian domain of the size 
$(L_x, L_y, L_z )=(192,192,64)\,r_j$ for models with $\kappa=0.5,\,1.0$ and 1.5 and 
$(1536,1536,64)\,r_j$ for the model with $\kappa=2.0$. The jet is centered on the $z$ axis.

\subsection{Results}
\label{sec:results}

As expected,  models with higher $\kappa$ turned out to be more stable. Here we first 
describe our naked eye observations and then provide with quantitative analysis. 
    
In the models with $\kappa=0.0$ and 0.5, which are relevant for jets surrounded 
by cocoons (radio lobes in AGN jets),  
the instability leads to complete disintegration of jets by the end of the 
simulation runs. 
Their evolution proceeds in a very similar way but takes 2-3 times longer for 
the $\kappa=0.5$ model.  The time-evolution of the $\kappa=0.5$ run is 
illustrated in figure \ref{fig:tevol_k0-5}.  
At t=100, where the corresponding non-expanding jet is already disrupted, the
$\kappa=0.5$ model has just entered he non-linear phase. The lateral jet protrusions 
drive weak compression waves into the ambient medium.  The deformations of the jet
column become comparable to the jet radius around $t\simeq200$.  
At this stage, the compression waves have turned into strong shocks, which begin  
to transfer a significant amount of the jet power to the environment. 
By $t=300$ the jet has lost its integrity and its fragments drive strong 
bow-shocks.  
By $t=500$, the effective jet radius has increased
to $\approx50 r_{\rm j}$, while the size of the corresponding
steady-state solution is only $\approx 4 r_{\rm j}$ at this time.  
Due to this dramatic increase in the jet cross
section and mixing with the ambient medium, the flow velocity has dropped to
a sub-relativistic level. The jet has now been totally destroyed and turned
into a turbulent plume.  

\begin{figure}
\includegraphics[width=8cm]{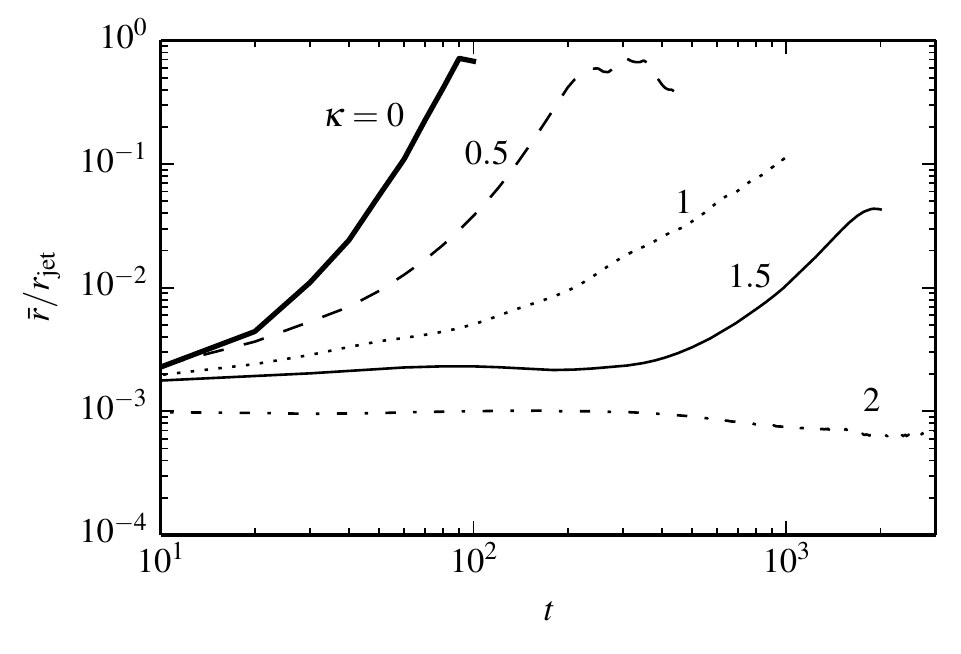}
\caption{Barycenter displacement normalized to the instantaneous jet
  radius for increasing values of $\kappa$.  When $\bar{r}/r_{\rm
    jet}\gtrsim0.5$, the jet looses integrity and disrupts.}
\label{fig:rcm}
\end{figure}

In the model with $\kappa=1.0$, the value at the lower end for 
galactic coronas, the jet shows significant fragmentation only 
by the end of the run, at $t=1000$. As one can see in figure~\ref{fig:3}, 
the jet core has fragmented into fast ``bullets'' that move through slower 
and less deformed envelope. The bullet's Lorentz factor is higher compared to 
the initial one of the core by a factor of two -- the result of prior thermal 
acceleration.   

In the model with $\kappa=1.5$, the value on the higher end for galactic coronas, 
the separation of the core and the envelope becomes even more prominent 
(see figure~\ref{fig:2}). By the end of the run ($t=2000$), the jet radius  
is approximately the same as in the corresponding steady-state solution and 
its envelope does not show noticeable deformations. However, already at $t=1000$ the jet core 
shows wiggles that have grown out of the initial $n=4$ mode of the perturbation.    
These deformations, advected with the fast flow in
the core, drive compression waves reminiscent of bow shocks into the
jet envelope. One can trace each such wave to a particular wiggle of
the core. By $t=2000$, the initial perturbation starts to fragment the jet core.

Continuing the general trend, at the critical value of $\kappa=2$ the envelope shows no 
visible features. The core, however, begins to show noticeable wiggles at $t=3000$ 
(see figure~\ref{fig:kappa2}).

\begin{figure}
\includegraphics[width=8cm]{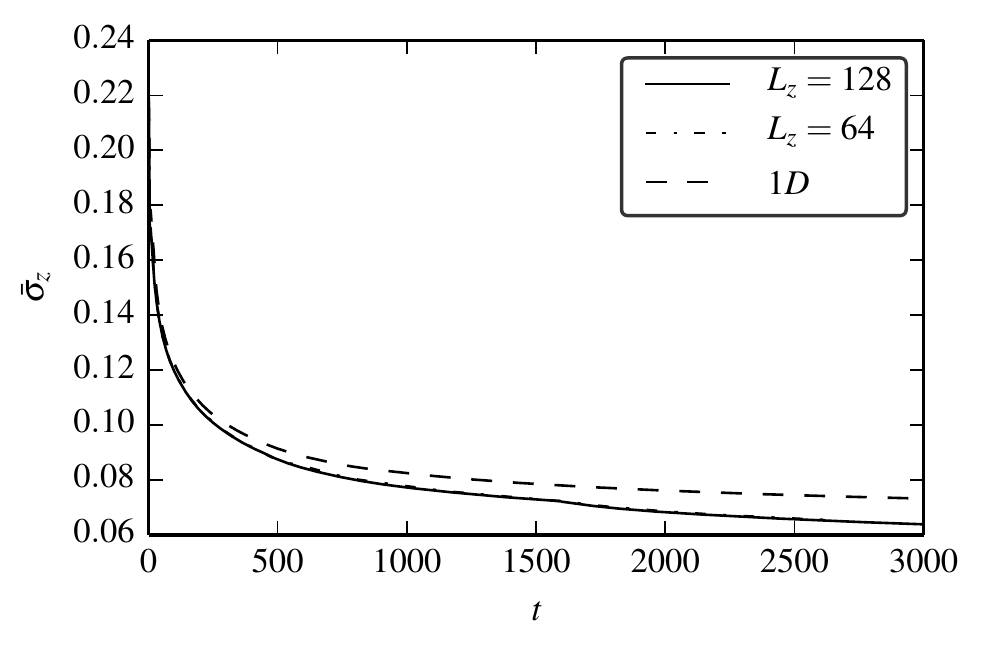}
\caption{Average jet magnetization in the $\kappa=2$ models comparing different
box sizes with the 1D case.  }
\label{fig:double}
\end{figure}

The growth of initial perturbations is reflected in the displacement of the 
jet center of mass, or barycenter. This displacement is a useful quantitative measure 
of the amplitude of global instability.  We compute the average barycenter 
displacement in the computational box via integration over the jet cross-section,
\beq
  \bar{r}=\frac{\left|\int Q \mathbf{r} ds \right|}{\int Q ds},
\eeq
and subsequent averaging along the jet axis for the whole box.  
As a weighting function we choose the relativistic inertia multiplied by the jet tracer
\beq
 Q=\Gamma^2 (\rho c^2+4 p) \tau \,. 
\eeq

Figure~\ref{fig:rcm} shows the evolution of $\bar{r}$ normalized to the current jet 
radius $r_{\rm jet}$ for models with different $\kappa$. 
Only in the runs with $\kappa=0$ and $0.5$ jets loose integrity and
become fully turbulent by the end of simulations. Based on these
models, we conclude that the disintegration occurs when
$\bar{r}=0.5r_{jet}$.  If the instability does not saturate prior to
reaching this amplitude, we expect the jet in the $\kappa=1$ model to
disintegrate around $t=\mbox{few}\times10^3$.  In the $\kappa=1.5$
model, we observe saturation of the core-instability at $t\simeq 2000$
and the jet does not loose global integrity due to the modes permitted
by the simulation.

Interestingly, for $\kappa=2$ the normalized barycenter displacement
is actually decreasing after $t\simeq400$, indicating that this jet
will never disintegrate, which is fully consistent with our theory.
In this run, the jet radius eventually exceeds the length of the
computational box, that prompts the question whether this can make a
strong impact on the simulation outcome. In order to investigate this
issue, we made another run with doubled $L_z$ dimension. For a fair
comparison, both cases were perturbed with the same vertical modes.
As evidenced in figure \ref{fig:double}, the resulting dynamics is
nearly indistinguishable, showing that the effect of the box size can
be neglected.

\begin{figure}
\includegraphics[width=8cm]{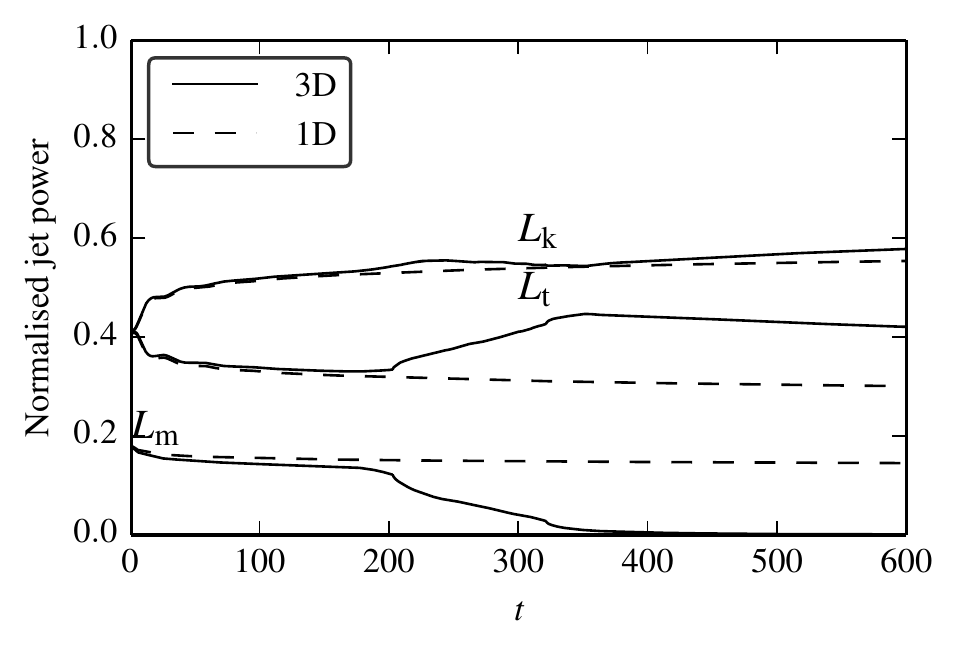}
\includegraphics[width=8cm]{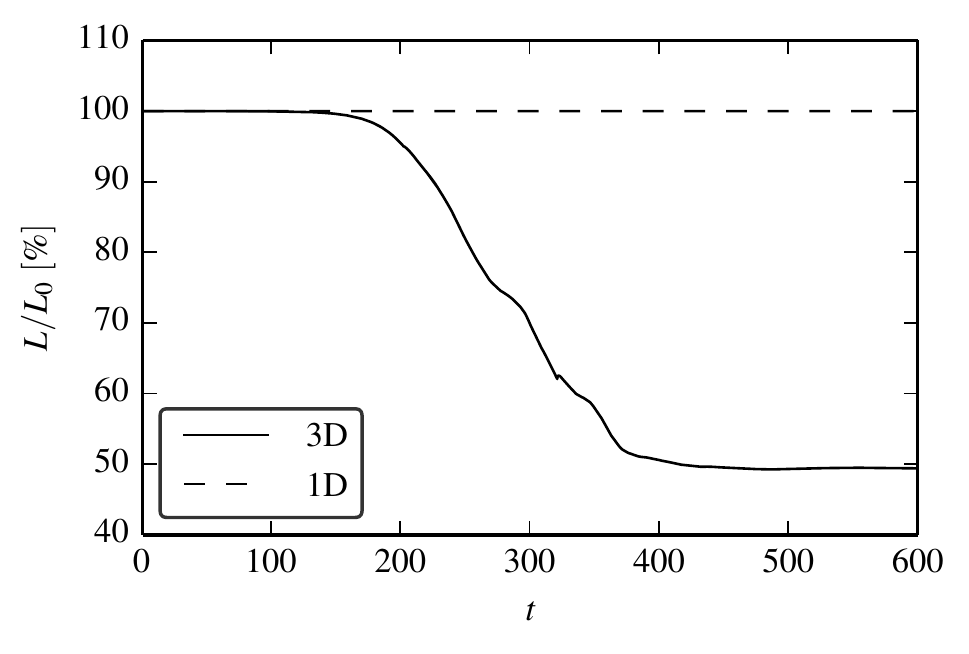}
\includegraphics[width=8cm]{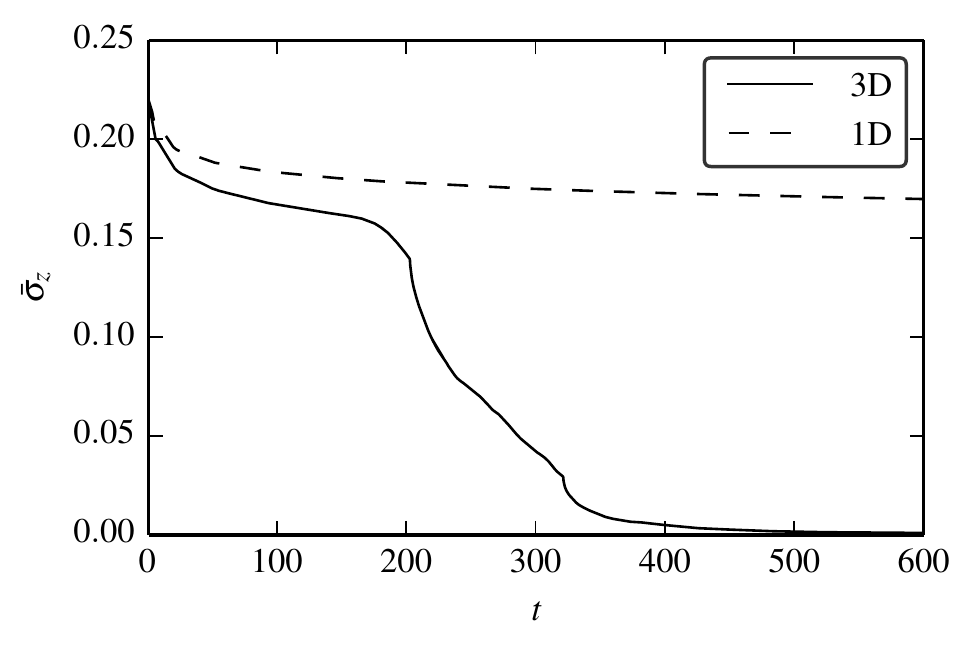}
\caption{Energetics of jets in the model with $\kappa=0.5$ -- 1D versus 3D.    
\textit{Left:} The contributions of kinetic, thermal and magnetic energy to the 
instantaneous total jet power. \textit{Middle:} Total jet power  
\textit{Right:} The volume averaged magnetization parameter $\sigma$. 
Dashed lines show the results for steady-state jets (1D simulations) whereas 
solid line show the results of time-dependent 3D simulations.   
 }
\label{fig:energetics}
\end{figure}
\begin{figure}
\includegraphics[width=8cm]{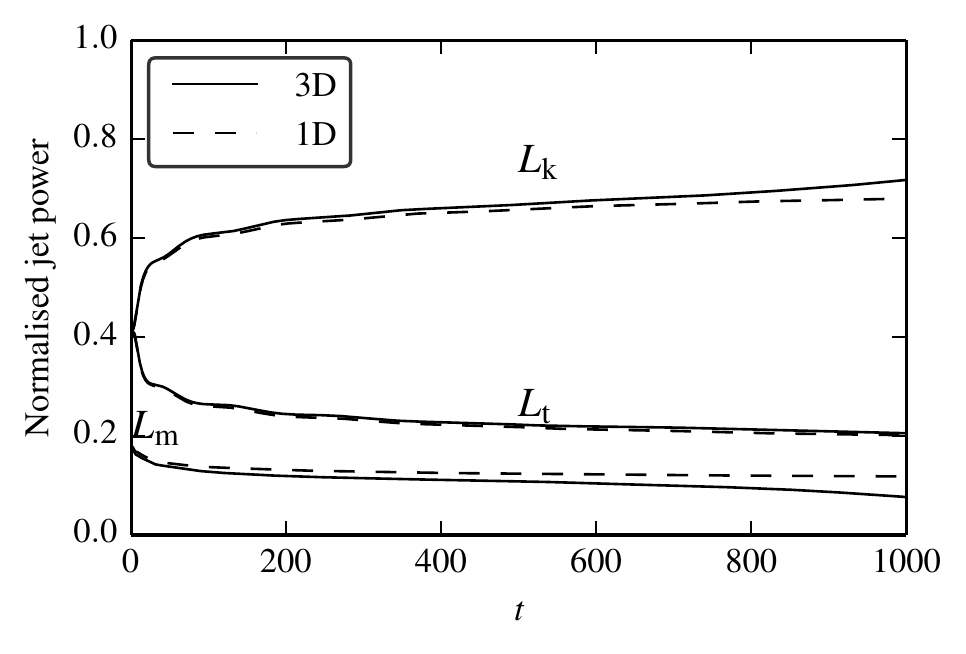}
\includegraphics[width=8cm]{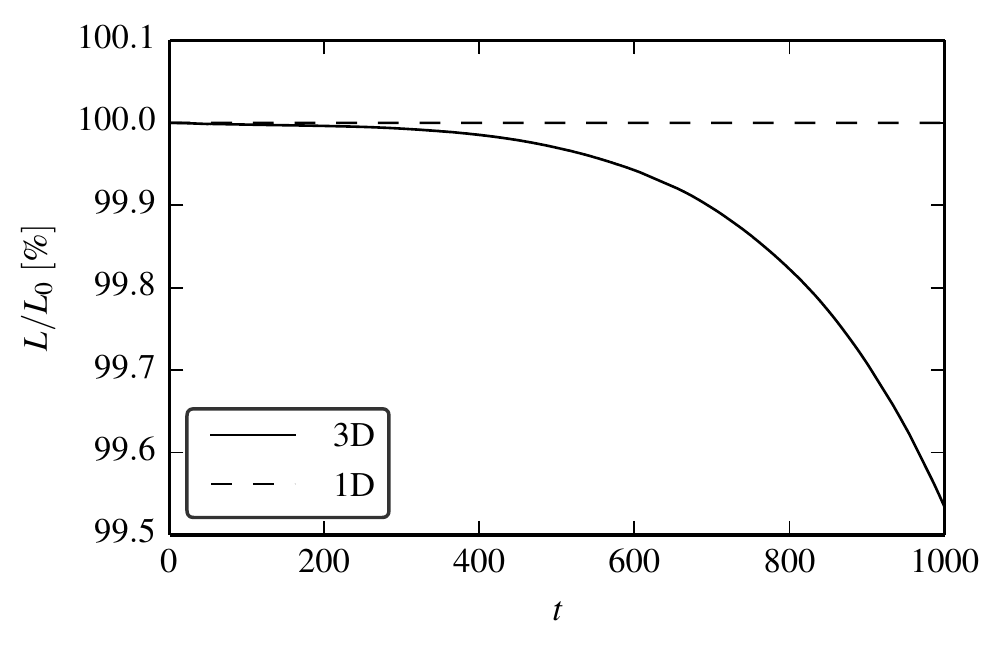}
\includegraphics[width=8cm]{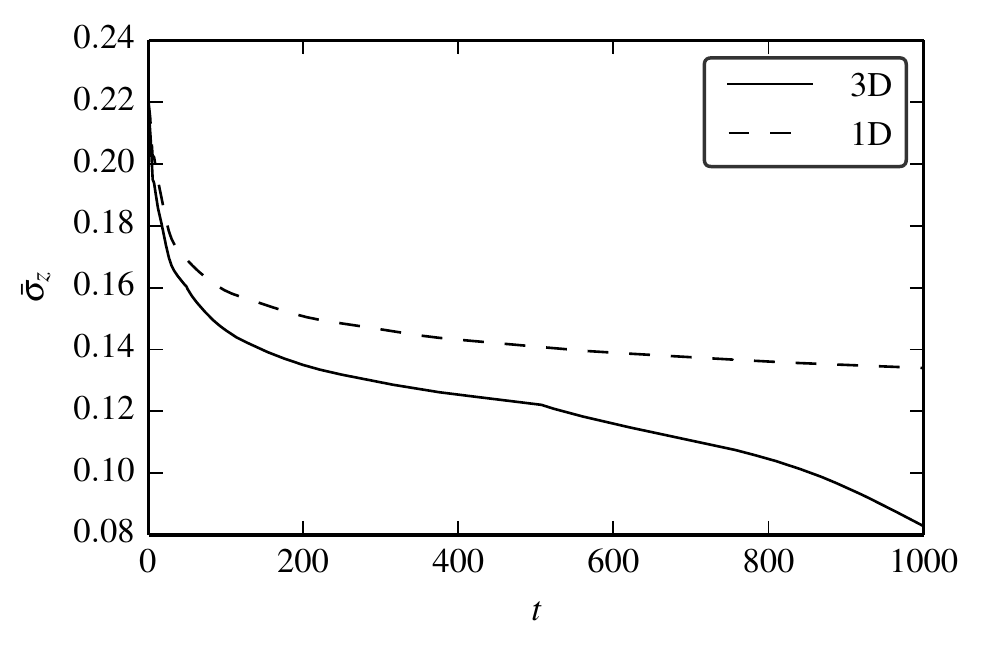}
\caption{As figure \ref{fig:energetics}, but for the model with $\kappa=1$.
 }
\label{fig:energetics_k1}
\end{figure}

\begin{figure*}
\begin{center}
\includegraphics[width=7.cm]{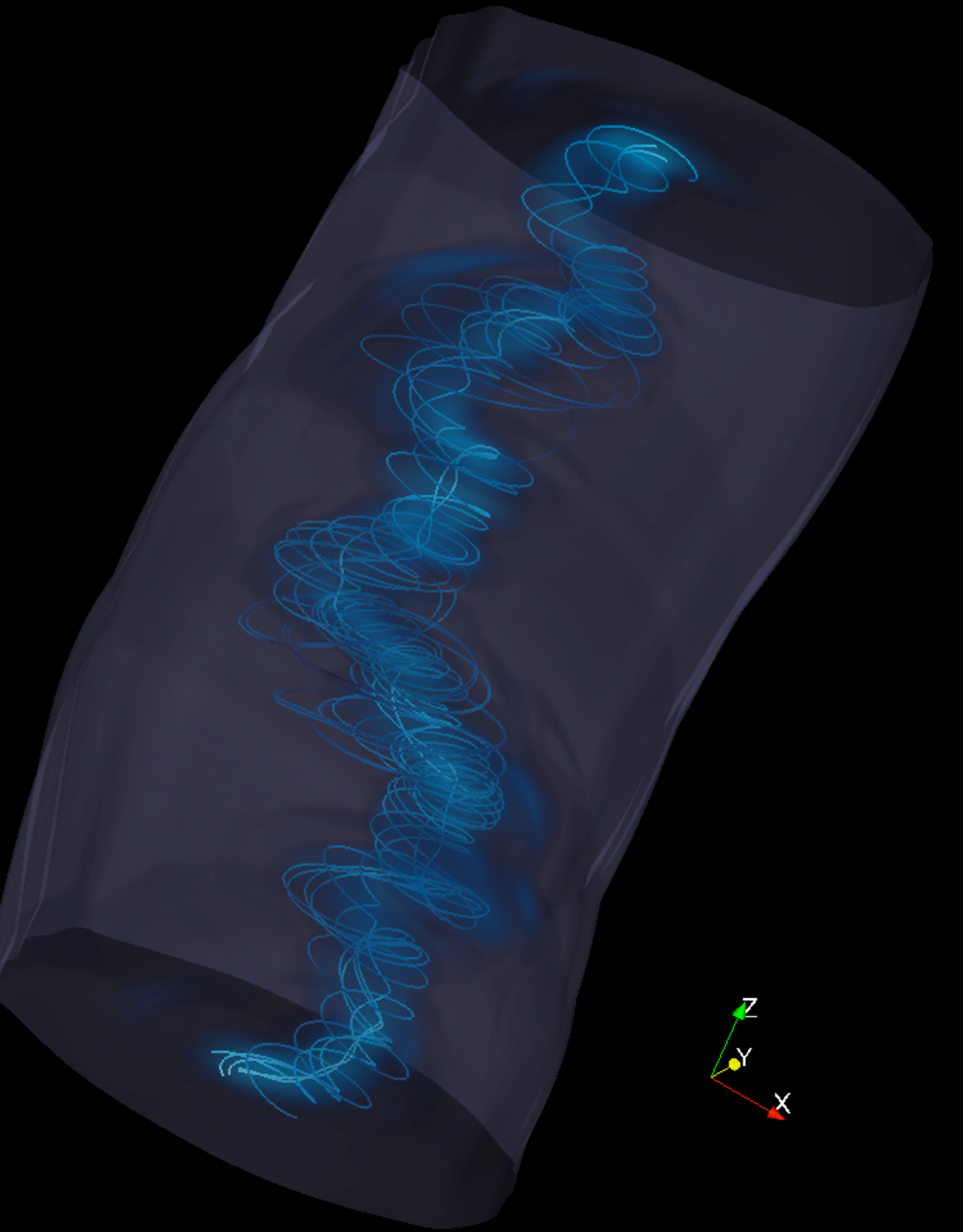}
\includegraphics[width=7.cm]{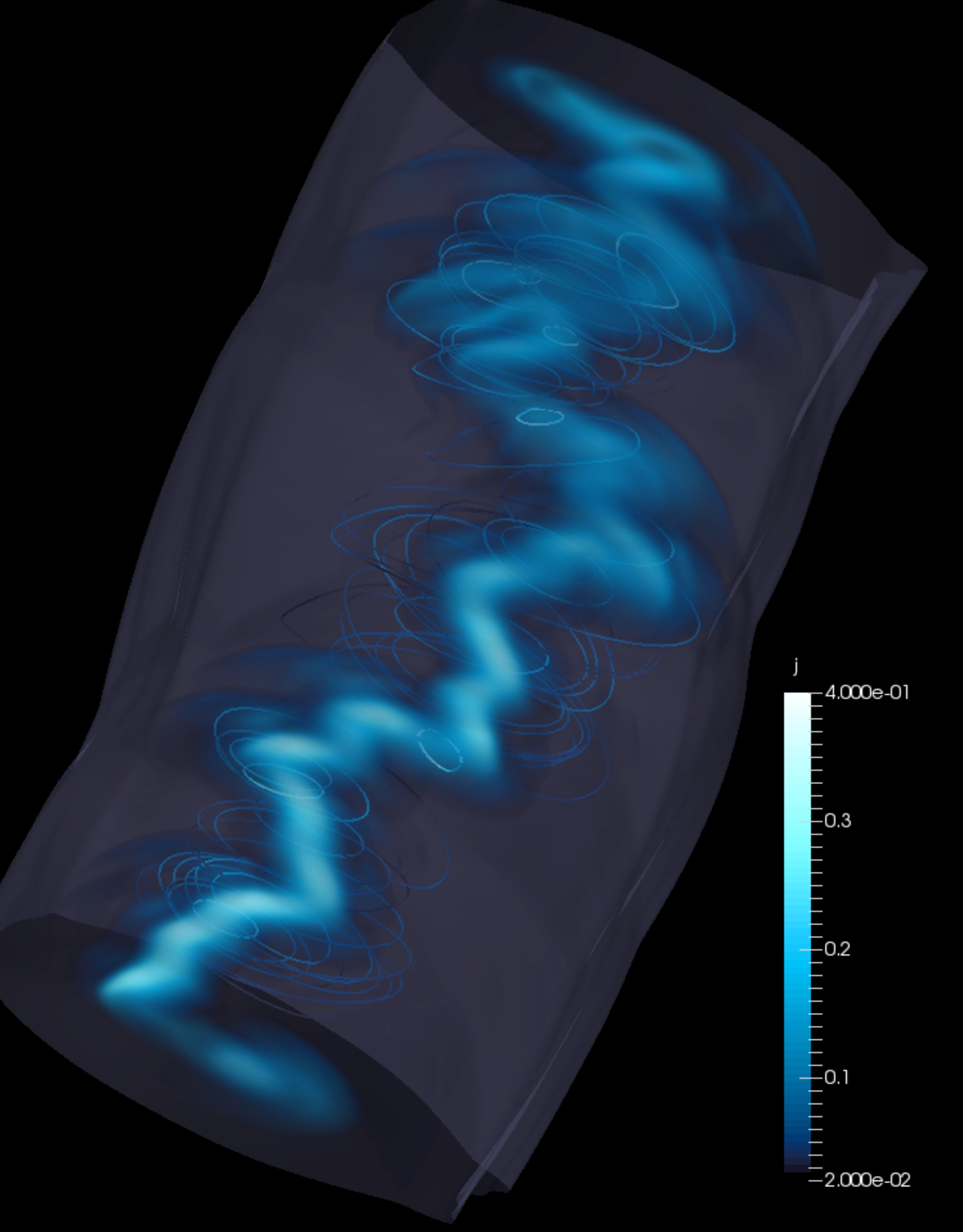}
\caption{ The electric current density and magnetic field structure in models 
with (left panel) and without (right panel) initial poloidal magnetic field 
($\kappa=1.0$ at $t=1000$). The outer contour corresponds to the jet boundary.  }
\label{fig:current}
\end{center}
\end{figure*}

Figure~\ref{fig:energetics} compares the energetics of our steady-state and 3D time-dependent 
solutions for the model with $\kappa=0.5$. One can see that initially, while the instability 
amplitude is still small, the jet evolution in both models is more or less the same. This is 
expected and only proves consistency between these two types of simulations.
The dominant process at this phase is the ideal MHD acceleration -- as the jet expands, its 
thermal and magnetic energy is converted into the kinetic energy of bulk motion. 
In fact, the mean magnetization parameter $\sigma$ decreases a bit faster in 3D, which is 
likely to be due to stronger magnetic dissipation\footnote{In the simulations we solve 
equations of ideal relativistic MHD and hence this dissipation is of numerical origin.}. 
The non-linear effects become important at $t\approx200$, where the difference between 
the two solutions becomes quite pronounced. 
The total of the 3D jet power decreases due to the energy transfer to 
the external gas via the shock waves driven at first by the jet wiggles and then by its 
fragments. The instability also generates current sheets inside the jet where the magnetic 
energy is dissipated. This is further illustrated in figure \ref{fig:current}, which shows 
a volume rendering of the current density for two runs with $\kappa=1$, 
one with initial  helical magnetic field and another with initial toroidal magnetic field.  
Both solutions exhibit snaking morphology that brings magnetic field lines of different 
directions closer together. In the model with pure toroidal initial field the current 
density is noticeably higher.  

As a result, the jet's Poynting flux rapidly decreases, and so does its mean 
magnetization. The dissipated magnetic energy is converted into heat. The process of magnetic 
dissipation develops rapidly and completes already at $z\approx 400$. After this point 
the total jet power remains more or less constant -- the jet is already very slow and does 
not drive strong waves into the external medium.

The energetics of the $\kappa=1$ model is consistent with the early phases of 
the $\kappa=0.5$ model (see figure~\ref{fig:energetics_k1}). 
In spite of displaying quite pronounced non-axisymmetric distortions and 
core fragmentation (see figure \ref{fig:2}), the jet of the $\kappa=1$ model 
looses only less than $1\%$ of its total power via emission of MHD waves 
by the end time of the simulation, at $t=1000$. 
For $\kappa=1.5$ and 2, the difference between the 
energetics of the steady-state and 3D models is even smaller.

\begin{figure*}
\begin{center}
\includegraphics[width=7.cm]{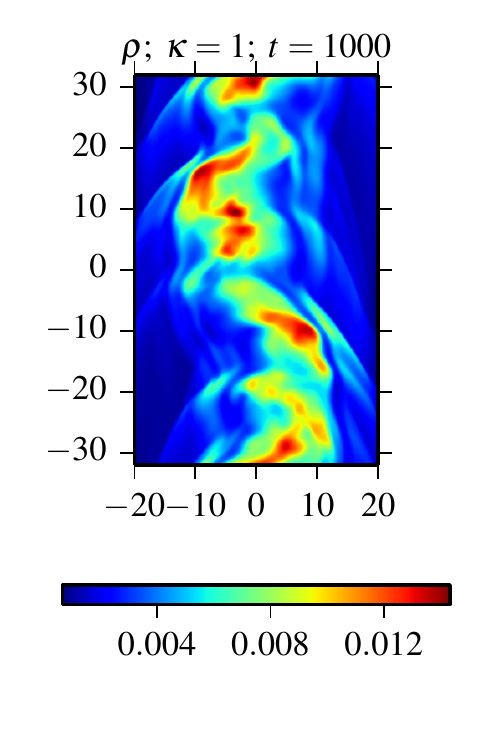}
\includegraphics[width=7.cm]{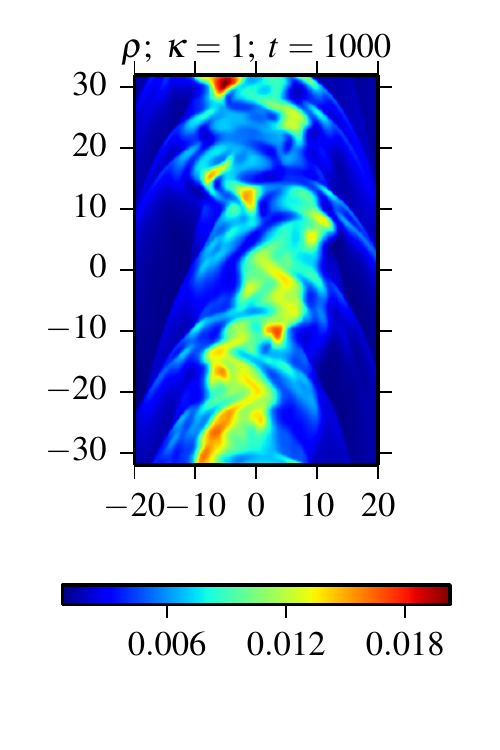}
\caption{Solutions for models with (left panel) and without (right panel) 
poloidal magnetic field. The plots show the distribution of rest mass density for models
with $\kappa=1.0$ at $t=1000$. }
\label{fig:4}
\end{center}
\end{figure*}

Cores of magnetically generated jets are likely to be dominated not by the gas pressure
but by the pressure of poloidal magnetic field. In order to explore the difference this can 
make on the jet stability, we made another run for the $\kappa=1.0$ model, now with 
poloidal magnetic field in the jet core.  In this model,
the initial radial profiles of the azimuthal component of the magnetic field, density and 
velocity are the same as before, but the gas pressure in the core is uniform and 
the magnetic field is initially force-free (see KPL for full details).  
As one can see in Figures~\ref{fig:current} and \ref{fig:4}, this modification has little 
effect on the jet stability -- in the non-linear regime, the morphology is very similar.  
However, we note that in the model with initial poloidal field the current density is 
higher and the current is less fragmented. This can be attributed to the role of the
magnetic tension associated with the poloidal field.

\section{Jet stability and the Fanaroff-Riley division} 
\label{sec:FR}

The issue of jet stability can be connected to the Fanaroff-Riley division of 
extragalactic radio sources into two basic morphological types \citep{FR-74} -- 
FR-I sources with wide two-sided kpc-scale jets and plume-like radio lobes (e.g. 3C~31)  
and FR-II sources with narrow one-sided jets and cocoon-like 
radio lobes with leading hot spots (e.g. Cygnus~A).  
In addition to the difference in  morphology, these two classes also differ in 
radio luminosity. The FR-II sources are not just more powerful on average, 
but there is a sharp division between the two classes on the radio-optical luminosity 
plane \citep{OL-94}. 

The structure of FR-II radio sources suggests that their jets survive all the way to the 
leading hot spots, where they collide with the surrounding medium \citep{BR-74}, whereas  
the structure of FR-I radio sources suggests that their jets suffer destructive global 
instabilities, become turbulent, entrain dense external gas, slow down to subsonic speeds 
and turn into plumes \citep{bicknell-84,Komissarov1990}. 
Observations indicate that on the pc-scale FR-I jets are similar to the FR-II jets --  
both could be described as well collimated relativistic outflows \citep{Venturi-95}.  
This indicates that the morphological difference between these types of jets on the kpc-scale 
is related to the different nature of their interaction with the environment on the 
length scales around 100 pc. 
 
Given the expected steep decline of the external gas pressure in the galactic nuclei,  
it is reasonable to assume that soon after leaving the immediate vicinity of the central 
black hole these jets become free and the global instabilities are suppressed. 
However, on the scale of about $100\,$pc they enter the region where the black hole gravity 
becomes small compared to that of the galaxy.  From the X-ray observations, we know that 
on this scale, the external gas pressure distribution flattens out - here the jets enter 
the central core region of the galactic X-ray coronas. Further out, on the scale of 
about 1 kiloparsec, the pressure begins to decline again, though not as steeply 
as inside AGN.    The pressure flattening in the coronal cores creates the necessary 
condition for the jet reconfinement. This can be important for the jet dynamics, as 
reconfined jets would becomes causally connected and hence susceptible to global 
instabilities. However, the reconfinement process can be too slow to be completed 
on the core scale. It involves a stationary ``conical'' shock wave (the reconfinement shock) 
gradually converging towards the jet axis. The rate of convergence depends, among other 
factors, on the jet power. The higher the jet power, the slower this rates becomes. 
This suggest that more powerful jets may fly through the galactic coronas unimpeded.     

The basic geometry of the reconfinement shock can be obtained in the Kompaneets 
approximation \citep{Komp-60} which for an unmagnetised relativistic jet 
leads to  
\beq
\oder{r}{z}-\frac{r}{z} = -z \fracp{p(z)}{K}^{1/2}  \,,   
\label{eq:rs1}
\eeq        
where $r$ is the shock radius, $K=\mu L_j/\pi\theta_0^2 c$, $L_j$ and $\theta_0$ is the 
power and opening angle of initially  free jet respectively and $\mu$ is a numerical constant
of order unity \citep{KF-97}. 
In terms of the shock opening angle $\theta=r/z$, dimensionless external pressure 
distribution $f(z)=p(z)/p_0$ and dimensionless distance $\zeta=z/z_0$, this equation 
reads  
\beq
\oder{\theta}{\zeta} = -\theta_0 \sqrt{\frac{f(\zeta)}{A}} \,,
\label{eq:rs2}
\eeq
where
\beq
    A=\frac{\mu L_j}{\pi p_0 z_0^2 c} 
\eeq 
is a dimensionless parameter that combines the effects of the jet power and ambient 
galactic pressure scale. Integrating Eq.(\ref{eq:rs2}), we find 
\beq
  \theta(\zeta)=\theta_0\left(1 - \int\limits_0^\zeta \sqrt{\frac{f(x)}{A}}dx  \right)\,. 
\eeq
which gives us the reconfinement scale  $\zeta_r$ via 
\beq
       \int\limits_0^{\zeta_r} \sqrt{f(x)}dx = \sqrt{A}\,.
\eeq
Interestingly, $\zeta_r$ does not depend on the jet opening angle. 

The pressure of galactic coronas is well represented by the model
\beq
    p=p_0 \left(1+(z/z_0)^2\right)^{-\kappa/2} 
\eeq
with $p_0\simeq 10^{-9}\mbox{dyn}\,\mbox{cm}^{-2}$, $z_0\simeq 1\,$kpc and 
$\kappa=1.25\pm0.25$ \citep[e.g.][]{MB-03}. Using these values we estimate 
\beq
   A\simeq\frac{1}{9} p_{0,-9}^{-1} L_{j,44} z_{0,\mbox{kpc}}^{-2}\, ,
\label{eq-A} 
\eeq
where $p_{0,-9}=p_0/10^{-9}\mbox{dyn}\,\mbox{cm}^{-2}$,  
$L_{j,44} = L_j/10^{44}\mbox{erg}\,\mbox{s}^{-1}$ and 
$z_{0,\mbox{kpc}} = z_0/1\,\mbox{kpc}$. Among these parameters, the most 
scattered one is the jet power and hence it is the jet power which mainly determines 
how far from the galactic center the jet becomes reconfined. 
Based on the energetics of cavities made by the jets in the hot gas of clusters 
of galaxies, \citet{CMN-10} give 
$10^{42}<L_j<10^{46}\mbox{erg}/\mbox{s}$, with $L_{FR}=10^{44}\mbox{erg}/\mbox{s}$ 
separating the FR-I and FR-II classes. 
Under the assumption that protons can be ignored in the jet energetics,  
\citet{GTG-09} derived similar top end jet powers via modelling the non€"thermal 
continuum of gamma-ray blazars. If, however, the jet plasma contains equal numbers 
of protons and electrons, the upper end of the power-range extends to $L_j = 10^{48}\mbox{erg}/\mbox{s}$.

\begin{figure}
\begin{center}
\includegraphics[width=6.cm]{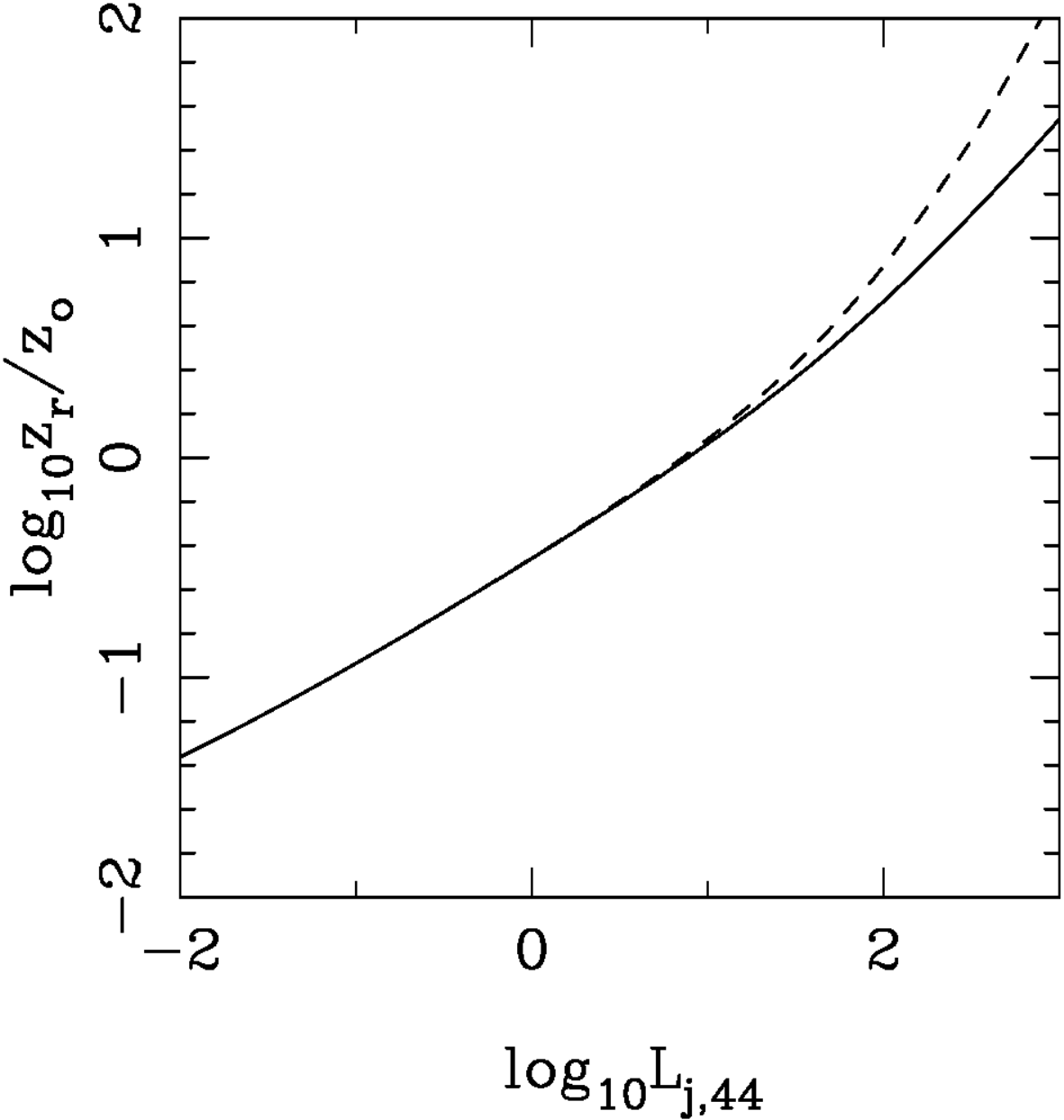}
\caption{Reconfinement scale in galactic corona as a function of the jet power
for $p_{0,-9}=1$ and $ z_{0,\mbox{kpc}}=1$. The solutions are shown for 
$\kappa =1$ (solid line)  and 1.5 (dashed line). 
}
\label{fig:5}
\end{center}
\end{figure}

Figure~\ref{fig:5} shows $z_r(L_{j})$ for fixed $p_{0,-9}=z_{0,\mbox{kpc}}=1$ and  
$\kappa = 1;\,1.5$, the lowest and the highest values of $\kappa$ which still 
agree with the observations \citep{MB-03}. 
One can see that for $L_j\ll L_{FR}$ the reconfinement occurs inside 
the coronal cores ( $z_r<z_0$) whereas for $L_j\gg L_{FR}$ well outside of them, at the 
distances more typical for the extended radio lobes. This result provides strong support
to the idea that the Fanaroff-Riley division is rooted in the jet stability.  
Namely, the jets with power $L_j < L_{FR}$ (FR-I jets) are reconfined inside the galactic 
coronas, where they become unstable and mix with the coronal plasma, whereas      
the jets with $L_j > L_{FR}$ (FR-II jets) remain unconfined and 
hence globally stable on the galactic scale.  The FR-II jets may get reconfined further
out, inside their radio lobes, where we expect a more-or-less uniform pressure distribution. 
The kink-mode instabilities in this region may be behind the observed wiggling of the FR-II
jets \citep[e.g][]{CB-96} and multiple hot spots \citep{Laing-81,Scheuer-82}. 
The non-destructive nature of these instabilities may be down to the 
fact that the jet mass density is higher than that of lobe density 
\citep[e.g.][]{2002ApJ...576..204H}.   

Another important aspect of the FR-division is the dependence of the critical radio 
luminosity, $P_{FR}$, on the optical luminosity of parent galaxies: 
$P_{FR}\propto L\sub{o}^{2}$  \citep{OL-94}. 
In our theory of the FR-division, this can only be explained via varied parameters of 
the interstellar gas distribution of parent galaxies \citep[cf.][]{Bicknell-95}. 
Equation~\ref{eq-A} shows that the critical jet power scales as $L_{FR}\propto p_0 z_0^2$, 
so it is important to know how $p_0$ and $z_0$ vary with the optical luminosity of 
the parent galaxy.

The stellar distribution of radio galaxies always shows the presence of a central core 
\citep{de-RuiterParma2005}. \citet{Kormendy-87} discovered that the core size increases with 
optical luminosity as $z_{0,*} \propto L\sub{o}^{1.1}$ 
\citep[see also][]{1997AJ....114.1771F}. The observations also reveal 
that the optical and X-ray surface brightness profiles of elliptical 
galaxies are almost identical \citep{TFC-86}, which suggests that the size of the X-ray 
core $z_0\propto L\sub{o}^{1.1}$ as well. The X-ray luminosity of the elliptical galaxies is 
approximately $L\sub{x}\propto L\sub{o}^2$ \citep{OFP-01}. Ignoring the weak dependence of the 
X-ray emissivity on the temperature, we have $L\sub{x}\propto n_0^2 z_0^3$, 
which yields $n_0\propto L\sub{o}^{-1/2}$. 
Assuming that the temperature itself arises from collisions of gas clouds ejected by 
stars, it depends on the stellar velocity dispersion as  $T\propto\sigma^2$  
The same result follows when the X-ray gas is modelled as an isothermal 
hydrostatic sphere \citep{Bicknell-95}. 
Given the Faber-Jackson relationship $\sigma \propto L\sub{o}^{0.25}$ 
\citep{FaberJackson1976,TDFB-81}, which is a convenient projection of the less scattered 
fundamental plane \citep{DresslerLynden-Bell1987,DjorgovskiDavis1987}, this reads as 
$T\propto L\sub{o}^{1/2}$ and hence $p_0 \propto n_0 T\propto L\sub{o}^0$ is independent 
of the optical luminosity. Thus we find that the critical jet power 

\be
    L_{FR} \propto p_0 z_0^2 \propto L_0^{2.2}\,, 
\ee   
which agrees very well with the observational results by \citet{OL-94}.

The Kompaneets approximation, which assumes constant pressure across the shocked 
layer, may lead to a substantial error for the reconfinement distance. For example, the 
calculations of \citet{NS-09}, which account for the pressure variation in the shocked 
layer, show a factor of two difference in the case of uniform external medium. 
Moreover, Eq.~\ref{eq:rs2} ignores the role of magnetic field. Future studies will 
clarify these issues.

\section{Discussion} 
\label{sec:discuss}

The results of our numerical simulations confirm the strong dependence of instabilities 
on the jet expansion associated with decline of external pressure. Although we 
considered only one particular case of transverse jet structure out of infinitely 
many possible configurations of various types, the underlying reason is very generic --
full or partial loss of causal connectivity -- and should operate for all types of 
supersonic jets, relativistic or not.   

In studies of cylindrical static columns, it has been found that our test 
configuration is disruptively unstable to kink mode instabilities, in contrast to 
configurations with force-free magnetic field, which are found to be much more stable 
and  exhibit only mild coherent deformations at the non-linear phase 
\citep{OBB-12}\footnote{In fact, they considered a somewhat different distribution of 
the azimuthal magnetic field in the envelope, with $b^\phi$ 
decreasing more like $r^{-2}$ than $r^{-1}$.  Nonetheless, we have seen that our 
cylindrical configuration was similarly unstable and hence this deviation is not 
significant.}.
Does this mean that such force-free configurations, where the magnetic force 
associated with the azimuthal component is finely balanced by the forces associated with
the poloidal component, are vastly superior and can provide an alternative explanation to 
the stability of cosmic jets?  Likely not. Indeed, as this has already 
been pointed out by \citet{OBB-12}, strong expansion of cosmic jets is bound to destroy such 
force-free equilibrium because in expanding jets the poloidal field decays faster than 
the azimuthal one.      
In magnetically dominated equilibrium jets this should lead to $b^\phi\propto r^{-1}$, 
which is exactly what we postulated for the envelope of our initial solutions. 
Our runs with and without the poloidal component already support this conclusion and 
future studies will clarify this issue further. 

On the one hand, mild global deformations are preferable because they preserve jet integrity. 
On the other hand, they are also deficient, being unable to trigger dissipation, 
particle acceleration, and ultimately non-thermal emission of cosmic jets. 
The solution to this conundrum could be found in local instabilities,       
which do not endanger the global integrity of jets.  In this regard, the core-envelope
structure of magnetised jets is a very attractive feature. Indeed, disintegration of the 
core via kink instability does not have to be fatal for the whole jet when the jet is much 
wider.

A slowly expanding core seems to be a generic property of jets with dynamically 
important helical magnetic field. For magnetically accelerated jets from rotating 
bodies this is well known \citep[e.g.][]{bog-95,BN-09}. In our models,  
there is no rotation and the slow expansion of the jet core is of a somewhat different nature. 
It arises from properties of the z-pinch equilibrium, where the hoop stress of 
the azimuthal component of magnetic field balances the pressure force.  
In order to elucidate this point, 
consider the non-relativistic case, which is much simpler. For magnetically-dominated 
envelope with predominantly azimuthal magnetic field, its force equilibrium requires  
$B_\phi=B_m (r/r_m)^{-1}$, in which case the hoop stress is balanced by the pressure 
force of the azimuthal field itself. However, this profile cannot be extended all 
the way to the jet axis, as this would lead to infinite magnetic energy. 
Instead it terminates at the core radius $r_m$.  In order to avoid infinite current 
density, $B_\phi$ should actually vanish at $r=0$. 
The conservation of magnetic flux of the envelope then requires
\be
   B_m \propto (r_m \log(r_{jet}/r_m))^{-1}\,.
\ee   
Inside the core, the hoop stress should be balanced either by the gas pressure or 
the pressure of the poloidal magnetic field.         
If the total core pressure is dominated by the gas contribution then it evolves 
as $p_c \propto r_m^{-2\gamma}$. The transverse force balance of the jet core can 
be approximated as $p_c/r_m \propto B_m^2/r_m$, which yields 
\be
   r_m\propto (\log (r_{jet}/r_m))^{1/(\gamma-1)} \,.
\label{eq:core}
\ee
If it is the pressure of the poloidal magnetic field which dominates in the core then 
Eq.~(\ref{eq:core}) still applies if we put  $\gamma=2$.  
One can see that in both these cases the core radius grows much slower than the jet radius. 
To put this in a different way, if the core radius expanded as fast as the jet radius 
than the hoop stress would decrease as $r_{jet}^{-3}$ whereas the core pressure 
force would drop faster, in conflict with the assumed force equilibrium.  

A fraction of the dissipated magnetic energy of the core can be emitted and the rest 
converted into the kinetic energy \citep[cf.][]{GS-06}. 
The result could be a fast and luminous spine surrounded by slow and relatively dark sheath. 
In the sheath, the magnetic field would be mainly
azimuthal, whereas in the spine its structure would be determined by the competition 
between the turbulent randomization, stretching by mean velocity field and shock compression.  
Such ``spine-sheath'' structure can explain many observations of AGN jets 
\citep[e.g.][]{Komissarov1990a,bridle-96,GDSW-07,GG-04,SO-02,GTG-05,LB-14}. Fragmentation 
of the jet core can also explain why the superluminal blobs of AGN jets occupy only a portion 
of the entire jet cross section \citep{lister-13}. The envelope, with its largely undisturbed 
azimuthal field, can be behind the transverse gradients of the Faraday rotation measure 
found in many parsec-scale AGN jets \citep{gabuzda-13}.   If the whole jet was affected by 
the kink instability and developed turbulence, one would not expect to find regular 
azimuthal field and hence the Faraday-rotation gradients on the scale of jet radius.    
 
VLBI observations of AGN jets allow to measure the Lorentz factor of their moving 
blobs $\Gamma_b$ and the jet half-opening angle $\theta_j$. These measurements lead to the 
result $\Gamma_b\theta_j \approx 0.2$ \citep{Jorstad-05,C-B-13}, which seems to indicate that 
these jets are causally connected and therefore externally confined. This conclusion 
is supported by the observed acceleration of the blobs in blazar jets up to the 
de-projected distance of $\simeq 100\,$pc \citep{mojave-xii}, which turns into a 
deceleration at larger distances. Indeed, this can been interpreted as an evidence 
of the magnetic collimation acceleration mechanism \citep{vk-04,kbvk-07,kvkb-09,lyub-10}, 
which requires external confinement. 
In fact, the observations of the M87 jet are not only in agreement  
with the MOJAVE data on the continued acceleration up to the distance of 
$\simeq 100\,$pc, where the stationary HST-1 feature is located, but also indicate 
the parabolic shape of the jet, $z\propto r^{1.7}$, 
in this region \citep{AN-12,Asada-14}. 
Further out the jet becomes conical and its speed decreases. These data challenge the 
key point of our theory that on the scales below 100pc the AGN jets are unconfined.

However, a closer look at the observational data shows a number of problems with 
this interpretation. For $p\propto z^{-\kappa}$ with $\kappa<2$ the collimation 
acceleration theory predicts the jet shape $z\propto r^a$, where the index 
$a=4/\kappa>2$ \citep{kvkb-09,lyub-09}, in conflict with  the observations of M87 
which give $a<2$. \citet{AN-12} applied the theoretical results for the degenerate case 
$\kappa=2$, which allows $1<a<2$ depending on the initial jet structure. However, 
because these transitional cases require $\kappa$ to be exactly two, it is hard to see
how they can be of more than just mathematical interest. For $\kappa>2$ the flows are 
asymptotically conical.

Moreover, for $\kappa<2$ the theory predicts $\Gamma_j \simeq r/r_{lc}$, where $r_{lc}$
is the radius of the light cylinder. For a Blandford-Znajek jet from a rapidly 
rotating black hole, $r_{lc} \approx 2r_s$, where $r_s=2GM/c^2$ is the Schwarzschild 
radius. Thus, one would expect the typical Lorentz factor of blazar jets $\Gamma\approx 10$
to be reached when the jet radius just exceeds $r_j=20 r_s$. In contrast, 
for the typical half-opening angle of blazar jets, $\theta_j=0.02$, the jet 
radius at the distance of 100pc is much larger:  
$r_j\approx 6\times 10^{18}\mbox{cm}\approx2\times10^4r_s$ for a $10^9M_\odot$ 
black hole. Thus, the asymptotic value of the Lorentz factor $\Gamma_{max}\approx10$ 
would have been reached on much smaller scales than observed. These estimates may not be very 
accurate for flows with only moderate asymptotic Lorentz factor, but the results of 
numerical simulations of such flows confirm that the acceleration of AGN jets should be 
almost fully completed inside the first parsec \citep{kbvk-07}.   

For a slowly rotating black hole $r_{ls}\approx (4/a)r_s$, where $a$ is the rotation parameter, 
so $\Gamma_j=10$ would be reached when $r_j=(40/a)r_s$. 
This can be matched with the observed jet radius 
at the end of the acceleration zone only for the incredibly low $a=0.002$.    
If the jet originates from a Keplerian disk then $r_{lc} = \sqrt{2}r_s (r_f/r_s)^{3/2}$,
where $r_f$ is the radius of the  magnetic foot-point on the disk. To match the observations, 
we would need $r_f\approx 140 r_s$, which is too far from the disk inner edge.    
In both these cases, it will be very difficult to explain the high power of blazar jets. 
 
It could be that the VLBI observations do not reveal the entire structure of AGN jets, but
only their bright magnetically confined inner cores. Their enhanced brightness is a 
combination of magnetic pinch and dissipation triggered by kink-mode instabilities. 
The magnetic dissipation may also power the observed bulk acceleration. It has been known
for some time that at the kpc scale even the minimal pressure of the  M87 jet is about 
one order of magnitude above the external gas pressure, which can be interpreted as an evidence 
of magnetic confinement \citep{BOH-83,OHC-89}. However, the polarization observations show that 
the magnetic field is predominantly aligned with the jet \citep{OHB-80,OHC-89}. 
This is qualitatively consistent with the M87 jet being only the jet core,  
where the magnetic field structure is randomized  by instabilities and stretched by 
velocity shear. It is well known that the poloidal field in kpc-jets cannot be regular 
as this leads to enormous magnetic flux, which cannot be sustained by any reasonable 
central engine \citep{BBR-84}. Where is the azimuthal magnetic field responsible 
for the magnetic confinement? Perhaps, it is in the free-expanding jet envelope 
which is dim because it is stable and the dissipation is not triggered in it and 
because it is slow and its emission is not Doppler-beamed. The Faraday rotation 
gradients across jets may have already revealed its presence \citep{gabuzda-13,AAN-13}.   

\citet{FW-85} explained the apparent over-pressuring of the M87 jet
by the compression at standing conical shocks. However, the pressure jumps in their
model are smaller than the observed ones, even if the minimal pressure represents the 
total jet pressure, and the structure of M87 knots does not have sharp features easily
associated with conical shock geometry \citep{BOH-83,OHC-89}.         

In this paper, focused primarily on AGN jets, which is a reflection of the authors
main research area.   However, the strong lateral expansion is a common property of all 
astrophysical jets and in this regard our results must have much broader application. 
Since these jets originate from central objects whose gravity dictates the properties 
of their environment, strong stratification with steep pressure gradients, 
promoting free expansion of jets, must be very common. Jets from young stars, X-ray binaries,
and collapsing stars are likely to be surrounded by broad winds originating in the 
same central objects. For a spherical adiabatic wind, the wind gas pressure drops
as $p\propto z^{-\kappa}$ with $\kappa = 2\gamma>2$. Adiabatic spherical accretion 
on a point-like central mass gives $\kappa=3\gamma/2$ \citep{Bondi}. 
In collapsing stellar envelopes, the gas pressure  follows a similar law, whereas for 
the ram pressure $\kappa=5/2$ \citep{Bethe-90}. In all these cases the pressure gradient 
is sufficiently steep to promote free expansion of jets and suppress instabilities via
the loss of causal connectivity. From the observations of ``jet-breaks'' in the emission 
of GRB afterglows one can estimate the product of the jet opening angle and its Lorentz
factor, $10<\Gamma_j\theta_j<50$ \citep{PanaitescuKumar2002}. For kinetic energy 
dominated flows this actually implies a total loss of causal connectivity 
\citep[e.g.][]{Zakamska2008}.  However, for Poynting-dominated flows the condition 
$\Gamma_j\theta_j>1$ is not sufficient to infer a causally disconnected flow as the 
fast-magnetosonic sound speed can be much closer to the speed of light \citep{kvkb-09}. 

The periodic box approach has its obvious limitations - it does not allow us 
to study wavelengths exceeding the box size and does not fully reproduce the 
conditions in expanding jets. In future, larger boxes may have to be
utilised for systematic studies of flows with strong poloidal magnetic field, which may 
suppress the growth of short-wavelength perturbations.  
Another option is to resort to computationally expensive simulations in
large non-periodic boxes. In this case, on can include
the magnetorotational central engine as a part of the problem and ensure
that the jet structure is consistent with its origin, which is an
indisputable advantage \citep[see e.g.][]{MSO-08,BM-09,Porth2013}.  
Interestingly, the stable jet simulated in \citet{BM-09} has an almost conical geometry, 
which is a characteristic of free expansion in steep ``atmosphere''. 
In these simulations the jet engine is initially surrounded by an almost empty space 
but later the inner region if filled with the disk wind. Overall, the pressure
decreases with distance faster than $z^{-2}$ (McKinney, private
communication), consistent with the conclusions of our work.

In this paper, we focused on relatively simple atmospheres described by power-law and King-type 
distributions. The reality is likely to be more complicated. For example, AGN jets may cross 
quasi-standing shocks resulting from the interaction between a wide disk wind and ISM. This would 
put the jet strongly off the lateral force balance with ISM after the crossing and drive in 
a reconfinement shock. In such strongly off-balance cases the reconfinement shock may actually 
reach the jet center even when $\kappa>2$ \citep{BL-07,KB-12}. A non-relativistic  magnetised 
disk wind may also play an important role in collimating the jets \citep{GTB-05}. We plan to 
explore these avenues in future studies.

\section{Conclusions}
\label{sec:concl}

Typical environmental conditions of cosmic jets include rapid decline of 
pressure with distance from the jet source. Our analysis shows that for atmospheres 
with the power law $p_{ext}\propto z^{-\kappa}$ pressure distribution, the value 
$\kappa=2$ is critical in the sense that a steeper pressure decline leads 
to such a rapid lateral jet expansion that the causal communication across the 
jet is completely lost and hence global instabilities of any type become totally 
suppressed. We propose that this is the reason for the observed remarkable stability 
of jets from young stars and AGN, which are capable of propagating distances which 
exceed their initial radius up to a billion times.         

Our numerical simulations are in full agreement with this conclusion. They 
convincingly demonstrate the reduction of the growth rate of the kink instability
with increase of the power index $\kappa$, and suggest global jet stability 
for $\kappa\ge 2$. In the simulations, we considered only one particular type of 
jets, but in combination with the very general analytical arguments they make a 
strong case in favour of the proposed explanation of the apparent stability of 
cosmic jets.      

When cosmic jets enter flat sections of external atmospheres, they may
re-confine and re-establish causal connectivity. This creates conditions for 
global instability. We have analysed the reconfinement process of extragalactic 
jets in the X-ray coronas of their parent galaxies, which have relatively flat
pressure distribution, and found that, depending on the jet power, the reconfinement
may occur both deeply inside the coronal core and well outside of it, on scales 
more characteristic of radio lobes. The separation between the two cases roughly corresponds to 
the jet power at the border line between FR-I and FR-II radio sources. This suggests
that the FR-I jets get re-confined, become unstable, and form turbulent plumes on the 
scale of the coronal core, whereas the FR-II jets burst through the corona 
largely unscathed. The critical jet power depends on the pressure and radius of the 
X-ray core. Using the empirical properties of elliptical galaxies, we derived the 
relationship between critical power and the optical luminosity of the host galaxy,
which in a very good agreement with the observations. 

Jets with dynamically-important magnetic field tend to be highly non-uniform, owing to 
the hoop stress of the azimuthal component of the magnetic field. 
When a jet develops a z-pinched core, this core expands much slower than the jet 
envelope and can preserve causal connectivity across itself. As the result, it becomes 
susceptible to instabilities. In our simulations we observed non-linear development of 
such instabilities, which resulted in core fragmentation and its energy dissipation. Such 
local instabilities do not present a threat to the integrity of the whole jet but they may be 
responsible for its observed emission and morphology. The so-called ``spine-sheath'' 
structure of AGN, supported by various observations, is one likely outcome.

\section{Acknowledgments}
SSK and OP are supported by STFC under the standard grant
ST/I001816/1. The computations were carried out on the UK MHD cluster Arc-I in Leeds and 
Dirac-II in Durham.    
OP likes to thank Purdue University for kind hospitality.

\bibliographystyle{mn2e}

\bibliography{BibFiles/mypapers,BibFiles/lyubarsky,BibFiles/lyutikov,BibFiles/jets,BibFiles/hea,BibFiles/numerics,BibFiles/mix,BibFiles/astro}

\end{document}